\documentclass[prb]{revtex4-2} 

\usepackage{amsmath}
\usepackage{bm}
\usepackage{amsfonts}
\usepackage{graphicx}
\usepackage{caption}
\usepackage{subcaption}
\usepackage{empheq} 
\usepackage{color}

\usepackage[utf8]{inputenc}
\usepackage[T1]{fontenc}

\newtheorem{hipotese}{{\bf Hypothesis}}
\newtheorem{definicao}{{\bf Definition}}
\newtheorem{teorema}{{\bf Theorem}}
\newtheorem{renorm}{{\bf Renormalization}}
\newtheorem{observacao}{{\bf Observation}}

\begin{document}

\title{On the probability distributions of the force and potential energy for a system with an infinite number of random point sources\\}
\author{E. L. S. Silva}
\email{elssufpa@gmail.com}
 \affiliation{
Instituto de Física, Universidade de Brasília, CP 04455,
70919-970 Brasília, Brazil}
\author{L.H. Miranda-Filho}
 \email{lucmiranda@gmail.com}
\affiliation{Institute of Theoretical Physics, Chinese Academy of Sciences, Haidian, Beijing 100190, China}
\author{A. Figueiredo}
\email{annibalfig@gmail.com}
\affiliation{Instituto de Física and International Center for Condensed Matter Physics, Universidade de Brasília, CP 04455, 70919-970 Brasília, Brazil}
\date{\today}
.
\begin{abstract}
In this work, we study the probability distribution for the force and potential energy of a test particle interacting with $N$ point random sources in the limit $N\rightarrow\infty$.  The interaction is given by a central potential $V(R)=k/R^{\delta-1}$  in a $ d$-dimensional euclidean space, where $R$ is the random relative distance between the source and the test particle, $\delta$ is the force exponent, and $k$ is the coupling parameter.
In order to assure a well-defined limit for the probability distribution of the force and potential energy, we { must} renormalize the  coupling  parameter  and/or  the system size as a function of the number $N$ of sources.

We show the existence of three non-singular limits, depending on the exponent $\delta$ and the spatial dimension $d$.
(i) For $\delta<d$ the force and potential energy { converge} to their respective mean values. This limit is called Mean Field Limit.
(ii) For $\delta>d+1$ the potential energy  converges  to a random variable and the  force to a  random vector.  This limit is called Thermodynamic Limit. 
(iii) For $d<\delta<d+1$ the potential energy converges to its mean and the force to a random vector. This limit is called Mixed Limit

Also, we show the existence of  two singular limits: (iv) for $\delta=d$ the potential energy converges to its mean and the force to zero, and (v) for $\delta=d+1$  the energy converges to a finite value and the force to a random vector.
\end{abstract}

\maketitle

\section{Introduction}
The development of modern probability theory took place mainly between the 1920s and 1940s as a result of the efforts of many mathematicians, among whom we can highlight P. Lévy and A. Ya. Khintchine.
In particular, the study of stable laws was first developed by P. Lévy in 1923 \cite{Lévy-1924, Lévy-1923}.
Two years later, Khintchine and Kolmogorov \cite{Khintchine-Kolmogorov1925} started a study on the convergence of infinite series of mutually independent random variables and showed that the convergence of means and variances lead to almost certain convergence for the series.
This fundamental problem is related to the classic question about the Gaussian limit for the sum of independent random variables and the { Central Limit Theorem (C.L.T.)}.
In fact, the necessary and sufficient conditions for the convergence of these sums to Gaussian distributions were independently formulated by Khintchine \cite{Khintchine1935}, Lévy \cite{Lévy1936, Lévy1936-2} and W. Feller \cite{Feller1935}.

The fundamental role of infinitely divisible distributions in probability theory was first established by B. de Finetti in a set of pioneering works in the late 1920s \cite{Finetti1929, Finetti1929-2, Finetti1929-3, Finetti1930}.
Nevertheless, its first formal definition was given by Khintchine \cite{Khintchine1937-1}.
 The general version of this problem, known as generalized C.L.T., was investigated by Lévy in a set of works from 1934 to 1937 \cite{Lévy1934, Lévy1935, Lévy-book-1937}. 
This generalized version of the C.L.T. claims that the sum of a large number of independent and identically distributed (iid) random variables with infinite variance can converge, with appropriate scale renormalization, to a distribution that belongs to a special family known as stable distributions of Lévy.
Indeed, it is important to note that the pioneering studies on non-Gaussian stable laws were developed by Lévy in the early 1920s  \cite{Lévy-1924, Lévy-1923}.

Furthermore, in a 1937 paper, Khintchine \cite{Khintchine1937} showed that Lévy's approach can be seen as an extension of Kolmogorov's triangular scheme and because of this, nowadays, the canonical form of infinitely divisible distributions is known in the literature as the Lévy-Khintchine formula.
Soon after, Khintchine and Lévy \cite{Khintchine-Lévy1937} published a paper together in which they presented a theorem on the general representation formula for characteristic functions of stable distribution laws.

Since these results are related to the study of sums of $N$ random variables, they can be applied to describe the statistics of a field generated  by a system of $N$ random sources, that is, the field can be considered as the sum of $N$ random variables. In this context the C.L.T is extensively used for sums of independent and identically distributed random variables in classical \cite{Tsalis-2015, Tsalis-2010} and quantum systems \cite{Cushen-1971, Hepp-1973, Hepp2-1973, Ja-2010}.

The problem of a stochastic force due to $N$ point sources, randomly distributed over space and interacting with a test particle, has been addressed by the physics literature, especially for gravitational systems in the context of astrophysics. The basic pillars of this literature begin with the seminal article by Chandrasekhar \cite{Chandrasekhar1943}, from which mathematical methods are developed to calculate the probability distribution for the stochastic force. These works assume the hypothesis of uniformity and global isotropy for the spatial distribution of the sources.
For a good review and analysis of this problem, we recommend the articles by Padmanabhan \cite{Padmanabhan1990} and Chavanis \cite{Chavanis2009}. In particular, Chavanis' paper is the first to apply these methods to study more general problems of a non-gravitational central force in a $d$-dimensional space, where this force is assumed to be proportional to the inverse of a certain power $\delta \neq 2$ of the distance between two interacting particles.
Let us remark that the methods of this literature were developed later and independently of all the vast bibliography related to the Central Limit Theorem and, as far as we know, Chavanis \cite{Chavanis2009} was the first to propose that these results could be derived from the application of the Central Limit Theorem.

The suggestion contained in Chavanis' paper is developed in the reference \cite{Figueiredo2019}. This work shows that the stochastic force can be determined from the formulation of Lévy's Central Limit Theorem \cite{Lévy-1923} - for random variables - adapted for random vectors. In this article, it is shown that the Central Limit Theorem completely determines the possible renormalizations to guarantee the existence of a well-defined limit for the force probability distribution.
A detailed comparison with the results obtained by Chavanis is also made.
However, in the references \cite{Chavanis2009} \and \cite{Figueiredo2019}, it is not analyzed which of these renormalizations would also guarantee a well-defined probability distribution for the potential energy and its relationship with the force.

In this present work, we show that the Central Limit Theorem solves the problem that requires a well-defined limit for the probability distributions of both: the force and potential energy. In fact, from the various cases of renormalization obtained in the reference \cite{Figueiredo2019}, only five types of possible limits remain:
(i) the Mean Field Limit for $\delta<d$, (ii) the Thermodynamic Limit for $\delta>d+1$, (iii) the Mixed limit for $d<\delta<d+1$, (iv) the Singular Limit for $\delta=d$ and (v) the Singular Limit for $\delta=d+1$. Furthermore, it is shown that no assumptions about the uniformity and global isotropy of the sources are necessary.

The paper is organized as follows:
in section 2, we introduce some mathematical tools and the formulation of the Central Limit Theorem in terms of characteristic functions;
in section 3, we formulate the problem of a test particle submitted to a central field generated by $N$ point sources with random spatial positions. Also, we define the renormalization constants and the hypotheses in order to apply the Central Limit Theorem.
In section 4, one applies the Central Limit Theorem to obtain the limits for the force and potential energy. Moreover,  we analyze the renormalization conditions that ensure the existence of these limits.
In section 5, we analyze the results obtained in section 4 and discuss some of their consequences and physical meaning.
Finally, in section 6, we close the paper with some concluding remarks and perspectives on further development.

\section{Mathematical Background and The Central Limit Theorem}

A real random variable $V$ has a probability distribution characterized by a given { probability measure} denoted by $\rho_V(v)$ with $v\in\mathbb{R}$, { and from now on just called ``measure''}.
A $d$-dimensional real random vector ${\vec V}=(V_1,\ldots, V_d)$ is an ordered collection of $d$ real random variables and its corresponding { measure}, defined for a given vector ${\vec v}\in\mathbb{R}^d$, is denoted by $\rho_{ \vec V}({\vec v})$.
We denote random variables and random vectors with capital letters, while elements belonging to their domain are denoted by lowercase letters. It is worth noting that a real random variable $V$ can be considered as a one-dimensional random vector ${\vec V}$ since we make the identification: ${\vec V}=(V)\in\mathbb{R}^ 1=\mathbb{R}$.
In this work, we consider that $\mathbb{R}^d$ is an Euclidean space.

Consider a vector transformation ${\bf T}$ acting on $\mathbb{R}^d$ and defined as:
\begin{equation}
\begin{array}{cccl}
{\bf T}: & \mathbb{R}^d&\longmapsto &\mathbb{R}^d \\
 & {\vec v} & \rightarrow & {\vec t}={\bf T}({\vec v}).
\end{array}
\label{eq1}
\end{equation}
The random vector ${\vec T}={\bf T}({\vec V})$  is obtained by applying the transformation ${\bf T}$ { on} the random vector
${\vec V}$. We must understand ${\bf T}({\vec V})$ as a transformation between the { measures} $\rho_{\vec V}({\vec v})$ and $\rho_{\vec T}({\vec t})$ induced by the vector function ${\vec t}={\bf T}({\vec v})$.

Let us consider a transformation ${\bf T}$ that is injective on $\mathbb{R}^d$ and differentiable 
in a closed set ${\cal U}\subset\mathbb{R}^d$. Therefore,
if the { measure} $\rho_{\vec V}({\vec v})$ is a continuous function on ${\cal U}$, then
\begin{equation}
\label{eq2}
\rho_{\vec T}\left({\vec t}\right)=\rho_{\vec V}\left({\vec v}\right)\left|\left|\frac{\partial {\vec v}}{\partial {\vec t}}\right|\right|,\;\;
\rho_{\vec V}\left({\vec v}\right)=\rho_{\vec T}\left({\vec t}\right)\left|\left|\frac{\partial {\vec t}}{\partial {\vec v}}\right|\right|,\;\; {\vec v}\in {\cal U},\;\;{\vec t}\in {\bf T}({\cal U}),
\end{equation}
${\bf T}:{\cal U}\rightarrow{\bf T}({\cal U})$ is an invertible transformation on ${\cal U}$ and $\rho_{\vec T}\left({\vec t}\right)$ is a continuous function on ${\bf T}({\cal U})$.
Moreover, ${\partial {\vec t}}/{\partial {\vec v}}$  and $\partial {\vec v}/\partial {\vec t}$ denote respectively the Jacobians of ${\bf T}$ and ${\bf T}^{-1}$, and $||\;||$ denotes the determinant modulus of a square matrix.

The mean of a transformation ${\bf G}:\mathbb{R}^d\longmapsto\mathbb{R}^n$ for a given random vector ${\vec V}$ is  defined as:
\begin{equation}
\label{eq5}
\left<{\bf G}({\vec V})\right>=\int_{\mathbb{R}^d}{\bf G}({\vec v})\rho_{\vec V}(\vec v)d^d{\vec v}=
{\int_{0}^{\infty}|{\vec v}|^{d-1}\int_{S_d}{\bf G}(|\vec v|{\hat v})d|\vec v|d\Omega_d},\;\;{\hat v}=\frac{\vec v}{|\vec v|}\in S_d,
\end{equation}
where $S_d$ is the spherical surface of unit radius centered at the origin of $\mathbb{R}^d$ and $d\Omega_d$ is the infinitesimal element of solid angle.
The solid angle $\Omega^{tot}_d$ of $S_d$ is given by

\begin{equation}
\label{eq6}
{\Omega^{tot}_d}=\int_{S_d}d\Omega_d=\frac{2\pi^{d/2}}{\Gamma\left(d/2\right)}.
\end{equation}

Let us remark that { if the integral in (\ref{eq5}) is finite,} then $\left<{\bf G}({\vec V})\right>\in\mathbb{R}^n$.
In particular, the mean of a random vector ${\vec V}$  is defined  as
\begin{equation}
\label{eq3}
\left<{\vec V}\right>=\left<{\bf I}(\vec V)\right>=\int_{\mathbb{R}^n}{\vec v}\rho_{\vec V}({\vec v})d^d{\vec v},
\end{equation}
where ${\bf I}:\mathbb{R}^d\longmapsto\mathbb{R}^d$ is the identity transformation, that is, ${\bf I}(\vec v)=\vec v$ for all $\vec v\in\mathbb{R}^d$.

The covariance matrix ${\mathbb{M}}_{\vec V}=[M_{ij}]_{\{i,j=1\cdots d\}}$ of a random vector ${\vec V}$ has their elements defined as
\begin{equation}
\label{covariance}
M_{ij}=\langle V_i V_j\rangle=\int_{\mathbb{R}^n}v_i v_j\rho_{\vec V}({\vec v})d^d{\vec v},
\end{equation}
where $v_i$ ($i=1,\ldots, d$) are the Cartesian components of the vector ${\vec v}\in\mathbb{R}^d$.

It is worth to recall that the mean and the covariance matrix of a random vector ${\vec V}$ can diverge depending on the {{ measure}} $\rho_{\vec V}(\vec v)$.

The characteristic function of a random vector ${\vec V}$  is defined as:
\begin{equation}
\label{eq4}
\Psi_{\vec V}({\vec z})=\left<e^{\displaystyle{\rm i}{\vec V}\cdot {\vec z}}\right>
=\int_{\mathbb{R}^d}{e^{\displaystyle{\rm i}{\vec v}\cdot {\vec z}}\rho_{\vec V}(\vec v)d^d{\vec v}},
\end{equation}
where ${\vec z}\in \mathbb{R}^d$. Remember that the characteristic function is a well-defined complex function for any vector ${\vec z}$ and ${\rm i}=\sqrt{-1}$. 

A random vector can be characterized by its { measure} {or by its} characteristic function, { where the characteristic function}  has the advantage of being well-behaved: continuous and differentiable. Furthermore, { the characteristic function allows a simpler way for} a functional characterization of infinitely divisible random vectors. On the other hand, although { measures} have an intuitive meaning, they can present very complicated irregularities since the only requirement is to be integrable.
The reference \cite{Lévy1929} provides an interesting discussion about the advantage of using characteristic functions instead of probabilities measure.
It is worth noting that the concept of a random vector is introduced solely to facilitate the language, always implying that any statement about a random vector is a statement about its respective characteristic function and/or its measure of probabilities.

We say that a sequence of random vectors $\left[{\vec V}_N\right]_{N=1,\ldots,\infty}$ converges to a random vector ${\vec W}$ if the sequence of characteristics functions 
$\left[\Psi_{{\vec V}_N}(\vec z)\right]_{N=1,\ldots,\infty}$ converges to $\Psi_{\vec W}(\vec z)$. Mathematically, we define:
\begin{equation}
\label{continuidade_de_Lévy}
\lim_{N\rightarrow\infty}{{\vec V}_N}=\vec W \;\;\Leftrightarrow\;\;\lim_{N\rightarrow\infty}\Psi_{{\vec V}_N}(\vec z)=\Psi_{\vec W}(\vec z)\;\;\forall \vec z\,\in\mathbb{R}^d.
\end{equation}
From Lévy's continuity theorem, the convergence of a sequence of characteristic functions leads to the convergence of their respective density { measures}.
For more details about probability distributions determination by characteristic functions and Lévy's continuity theorem, see reference \cite{Lévy1952}.

For a given $a\in\mathbb{R}$ and $\vec b\in\mathbb {R}^d$, we say that the random vector $\vec W=\vec b+a\vec V$ is a renormalization of ${\vec V}$. According to well-known properties of characteristic functions, we have
\begin{equation}
\label{fc_renormalizacao}
\Psi_{\vec W}(\vec z)=e^{\displaystyle{\rm i} \vec z\cdot \vec b}\Psi_{\vec V}(a\vec z).
\end{equation}
The vector $\vec b$ is called renormalization vector and the number $a$ renormalization constant.

Two random vectors $\vec V$ and $\vec W$ are said to be identical if their respective characteristic functions are the same, and we denote this identity relation as $\vec V\equiv\vec W$. Two vectors ${\vec V}$ and $\vec W$ are independent if and only if
\begin{equation}
\label{independencia}
\Psi_{\vec V+\vec W}(\vec z)=\Psi_{\vec V}(\vec z)\Psi_{\vec W}(\vec z).
\end{equation}

A random vector $\vec S$ is said to be stable if and only if there exist real number $a$ such that
\begin{equation}
\label{fc_estabilidade}
\Psi_{{\vec S}}(a\vec z)\Psi_{{\vec S}}(a\vec z)=\Psi_{\vec S}(\vec z).
\end{equation}
On the issue about the representation of stable random variables by characteristic functions see references \cite{Lévy-1924,Lévy-1923,Khintchine-Lévy1937}. On the representation of stable random vectors see references \cite{Lévy1936-2, Lévy1969,Taqqu,Meerschaert2001}.

A random vector $\vec S$ is said to be infinitely divisible if for every integer $N$ its respective characteristic function can be factored into the product of $N$ characteristic functions, { which are not necessarily identical}.
Every stable random vector is infinitely divisible and, in this case, for every $N\in\mathbb{N}$ there are real number $a_N$ such that
\begin{equation}
\label{divisibilidade}
\Psi_{\vec S}(\vec z)=\Psi_{\vec S}(a_N\vec z)\Psi_{{\vec S}}(a_N\vec z)\cdots\Psi_{\vec S}(a_N\vec z)\;\;\forall\,\vec z\in\mathbb{R}^d.
\end{equation}
On the representation of infinitely divisible random variables by characteristic functions, see references \cite{Khintchine1937-1,Lévy1938,Lévy1939}. On the question of representing infinitely divisible random vectors, see reference \cite{Meerschaert2001} and references therein.

Let us  consider the renormalized sum of $N$ identical and independent vectors: 
\begin{equation}
\label{soma}
\vec S_N={\vec b}_N+\sum_{i=1}^Na_N\vec V_i\; (i=1,\ldots,N), \;\;{\vec V}_i\equiv\vec V.
\end{equation}
Then, the main result of the Central Limit Theorem is to state that, with appropriate conditions on the renormalization vector ${\vec b}_N$ and constant $a_N$, the sequence of vectors
${\vec S}_N$ converges to a stable random vector $\vec S$. Mathematically, we have
\begin{equation}
\label{tlc}
\lim_{N\rightarrow\infty}{\vec S}_N=\vec S\;\;\Leftrightarrow\;\;
{ \lim_{N\rightarrow\infty}}\Psi_{\vec S_N}(\vec z)=
\lim_{N\rightarrow\infty}\left(e^{\displaystyle \vec z\cdot \vec b_N}
\left[\Psi_{\vec V}(a_N\vec z)\right]^N\right)=\Psi_{\vec S}(\vec z).
\end{equation}  

In the next subsection, we present a version of the Central Limit Theorem with $\vec b_N=\vec 0$, then we are tacking into account only the renormalization constants $a_N$. In other words, we are considering only scale renormalizations for the random vectors ${\vec V}_i$.

\subsection{Central Limit Theorem}

\begin{definicao}
\label{def4}
For a given random vector ${\vec V}$, we define its Lévy exponent (denoted by $\alpha$) as the lowest order of the non-analytical term of the series expansion of the characteristic function $\Psi_{\vec V} ({\vec z})$ around ${\vec z}=\vec 0$. Remember that this expansion can always be written as
$$\Psi_{\vec V}({\vec z})=1+\phi({\vec z})+I_\alpha({\vec z}) +{ R({\vec z})},\ ;\;I_\alpha(\vec z)=|{\vec z}|^\alpha \theta_\alpha({\vec z}),$$
where $\alpha>0$, $\phi({\vec z})$ is { a polynomial function} with powers smaller than $\alpha$, the function $\theta_\alpha({\vec z})$ can be written as
$$\begin{array}{c} \theta_\alpha({\vec z})=\theta_\alpha(\hat z)\;\; {\rm if}\;\; \alpha \notin \;\mathbb{N}, \\
    \theta_\alpha({\vec z})=\theta^1_\alpha(\hat z)\ln|\vec z|+\theta^2_\alpha(\hat z)\;\;{\rm if}\;\; \alpha\in\mathbb{N}, \end{array}$$
and the function ${ R({\vec z})}$ is such that $\displaystyle \lim_{\vec z\rightarrow \vec 0}{ R({\vec z})}/{I_\alpha({\vec z})}=0$.

If there is no non-analytic term in the expansion of the characteristic function, then we define the Lévy exponent as $\alpha=\infty$.
\end{definicao}

The possibility of writing the characteristic function of a random vector $V$, as established in Definition \ref{def4}, implies that its logarithm can be written as follows:
\begin{equation}
\label{definicao4}
\ln \Psi_{\vec V}(\vec z)=
\left\{\begin{array}{cl}
I_{\alpha}({\vec z})+\Omega(\vec z) & \;\;{\rm if}\;\;0<\alpha\leq 1, \\
{\rm i} {\vec z}\cdot\left<\vec V\right>+ I_{\alpha}({\vec z})+\Omega(\vec z) & \;\;{\rm if}\;\;1<\alpha\leq 2, \\
\displaystyle {\rm i} {\vec z}\cdot\left<\vec V\right>+\frac{1}{2}\vec z \cdot\left(\mathbb{M}_{\vec V}\vec z\right)+\Omega(\vec z) & \;\;{\rm if}\;\;2<\alpha\leq\infty,
\end{array}
\right.
\end{equation}
where $\mathbb{M}_{\vec V}$ is the covariance matrix of  $\vec V$. The function $\Omega(\vec z)$ is continuous at $\vec z={\vec 0}$ and
\begin{equation}
\label{definicao4a}
\lim_{\vec z\rightarrow \vec 0}\frac{\Omega(\vec z)}{I_\alpha(\vec z)}=0\;\;{\rm if}\;\;0<\alpha\leq 2,
\;\;\;\;\lim_{\vec z\rightarrow\vec 0}\frac{\Omega(\vec z)}{\vec z\cdot\left(\mathbb{M}_{\vec V}\vec z\right)}=0\;\;{\rm if}\;\;2<\alpha\leq\infty.
\end{equation}

The fact that $\Psi_{\vec V}(\vec z)$ is a characteristic function imposes restrictions on the function $I_\alpha(\vec z)$  for $0<\alpha\leq 2$, i.e., functions $\theta_\alpha(\vec z)$ can not be arbitrary. On the general form of this function $I_\alpha({\vec z})$ see references 
\cite{Lévy-1924,Lévy-1923,Khintchine-Kolmogorov1925,Lévy1936,Khintchine1937,Lévy1934,
Lévy-book-1937,Khintchine1937,Khintchine-Lévy1937} for random variables and \cite{Lévy1936-2,Lévy1969,Taqqu,Meerschaert2001} for random vectors.

It is worth to make one observation about the { Lévy exponent} given in Definition \ref{def4}.

\begin{observacao}
\label{obs1}
A random vector with a finite covariance matrix has Lévy exponent $\alpha>2$, and a random vector with a finite mean has $\alpha> 1$. Generally speaking, for a given $\alpha>0$, we have $\left<|{\vec V}|^{\beta}\right><\infty$ ($\beta<\alpha$) and $\left <|{\vec V}|^{\beta}\right>=\infty$ ($\beta\geq\alpha$). If the characteristic function is analytic, having a well-defined Taylor series for all orders, the Lévy exponent is $\alpha=\infty$.
\end{observacao}

Now, let us consider the sum in eq. (\ref{soma}) with ${\vec b}_N=\vec 0$. The Theorem \ref{teo1} below is a Central Limit Theorem that establishes the conditions on the renormalization constants $a_N$ in order to assure the convergence of ${\vec S}_N$ to a stable random vector for $N\rightarrow\infty$.

\begin{teorema}
\label{teo1}
{ Consider} a random vector ${\vec V}$ with Lévy exponent $\alpha>0$. Let us consider a sequence of {nonzero} real numbers $a_N$ ($N\in\mathbb{N}$) such that $q=a_N/|a_N|$ is the same for all $N$. Furthermore, for a given $K>0$, let us consider in Table \ref{tabela1} the following conditions imposed on $a_N$:

\begin{table}[h]
\begin{center}
\begin{tabular}{ccccc|ccccc|ccccc}
\hline\hline
& $0<\alpha<1$ & & & & & & $\alpha=1$  & & & & & &  $1<\alpha<\infty$  &\\ \hline\hline
& $N|a_N|^\alpha=K$ & & & & & &  $-N|a_N|\ln |a_N|=K$ & & & & & & $N|a_N|=K$ & \\ &&&&&&&&&&&&&& \\
& $\displaystyle |a_N|=\left(\frac{K}{N}\right)^{1/\alpha}$ & & & & & & $\displaystyle |a_N|=h\left(\frac{K}{N}\right)$ & & & & & & $\displaystyle |a_N|=\frac{K}{N}$ & \\   &&&&&&&&&&&&&& \\
& $N>0$  & & & & & &  $\displaystyle N\geq e K$ & & & & & & $N>0$ & \\ 
\hline\hline
\end{tabular}
\end{center}
\vspace*{-5mm}
\caption{The first line represents the renormalization condition for the respective range of $\alpha$, the second line represents the solution for this condition and the third line stand for the range of $N$ for which this condition has a solution. The function $h(x)$ is the solution of equation $-h(x)\ln h(x)=x$ for $x\in[0,e^{-1}]$.}
\label{tabela1}
\end{table}

Then, for $0<\alpha\leq 2$, there are well defined functions $A_\alpha(\hat z)$, $B_\alpha(\hat z)$ for all $\hat z\in S_d$, for $\alpha>1$ there is a well defined mean $\left<\vec V\right>$ and for $\alpha>2$ there is a well defined covariance matrix $\mathbb{M}_{\vec V}$ for the vector $\vec V$  such that { we have asymptotically for $N\rightarrow\infty$:}
$${ \left(\Psi_{a_N\vec{V}}(\vec z)\right)^N=\left(\Psi_{\vec{V}}(a_{N}\vec z)\right)^N\asymp \psi_\alpha(N,{\vec z}),}$$
where $\ln\psi_\alpha(N,{\vec z})$ { is given by}
\begin{equation}
\label{eq18}
\ln\psi_\alpha(N,{\vec z})=
\left\{\hspace*{2mm}
\begin{array}{ll}
\hspace*{-2mm}\displaystyle -\frac{K|\vec{z}|^\alpha}{\Gamma(\alpha +1)}G_\alpha(\hat z) & \;\;{\rm if}\;\; 0<\alpha<1, 
\\ \\
\hspace*{-2mm}\displaystyle-{\rm i}|\vec z|K B_1(q\hat z)- \frac{\pi \sigma_N|\vec z|}{2} A_1(\hat z) & \;\;{\rm if}\;\; \alpha=1, 
\\ \\
\hspace*{-2mm}\displaystyle{\rm i}\vec z\cdot\langle qK\vec V\rangle-\frac {\sigma_N^\alpha|\vec{z}|^\alpha }{\Gamma(\alpha +1)}G_\alpha(\hat z) &\;\;{\rm if}\;\;1<\alpha< 2,
\\ \\
\hspace*{-2mm}\displaystyle{\rm i}\vec z\cdot\langle qK\vec V\rangle-\frac {\sigma_N^2|\vec{z}|^\alpha }{4}A_2(\hat z) & \;\;{\rm if} \;\;{\rm if}\;\;\alpha= 2,
\\ \\
\hspace*{-2mm}\displaystyle{\rm i}\vec z\cdot \langle{qK\vec V}\rangle-\frac{\sigma_N^2}{2}\vec{z}\cdot\left(\mathbb{M}_{\vec V}\vec{z}\right) & \;\;{\rm if}\;\;(2<\alpha<\infty),
\end{array}\right.
\end{equation}
with $G_\alpha(\hat z)$ for $0<\alpha<2$ ($\alpha\neq 1$)  given by
$$
\displaystyle G_\alpha(\hat z)=\pi\frac{\cos(\alpha\pi/2)A_\alpha(\hat{z})-{\rm i}\sin(\alpha\pi/2)B_\alpha(q\hat{z})}{\sin(\alpha\pi)}.
$$
The corresponding expressions for $\sigma_N$, for the different {ranges} of $\alpha$, are given in Table \ref{tabela2}.
\vspace*{-3mm}
\begin{table}[h]
\begin{center}
\begin{tabular}{c}
\begin{tabular}{ccc|ccc|cccc|ccc}
\\ \hline\hline
$\alpha=1$ & & & & $1<\alpha<2$ & & & & $\alpha=2$ & & & & $2<\alpha\leq\infty$\\
$\sigma_N=Nh(K/N)$ & & & & $\sigma_N=KN^{(1-\alpha)/\alpha}$ & & & & $\sigma_N=K\left(N^{-1}\ln N\right)^{1/2}$ & & & &  $\displaystyle\sigma_N=\frac{KN^{-1/2}}{2}$ 
\\ \hline\hline
\end{tabular}
\end{tabular}
\end{center}
\vspace*{-5mm}
\caption{The expressions of $\sigma_N$ for the different {ranges} of $\alpha$.}
\label{tabela2}
\end{table}
Moreover, for $0<\alpha\leq 2$, the characteristic function $\psi_\alpha(N,{\vec z})$ corresponds to a stable Lévy probability distribution with exponent $\alpha$. For $\alpha > 2$, the characteristic function $\psi_\alpha(N,{\vec z})$ corresponds to a Gaussian probability distribution.

The function $\psi_\alpha(N,{\vec z})$ is just an asymptotic expression of the function $\left(\Psi_{\vec V}({a_N\vec z})\right)^N$, that is,
$$N\rightarrow\infty\;\;\Rightarrow\;\;\ln \psi_\alpha(N,{\vec z})\asymp N\ln \Psi_{\vec V}({a_N\vec z}).$$
The $\lim_{N\rightarrow\infty}\psi_\alpha(N,{\vec z})$ is obtained by tacking into account that $\lim_{N\rightarrow\infty}\sigma_N=0$, according with the expressions for $\sigma_N$ in the table above.
\end{teorema}
\vspace*{2mm}
\noindent{\it Proof}: The fundamental idea to { prove} this theorem is based on the property that the characteristic function of ${\vec V}$
can be written as in (\ref{definicao4}) and satisfies equation (\ref{definicao4a}).
Therefore, $\ln\left(\Psi_{\vec V}(a_N\vec z)\right)^N$ can be expressed as:
\begin{equation}
\label{new1}
N\ln\Psi_{\vec V}(a_N\vec z)=
N\times\left\{\begin{array}{cl}
I_{\alpha}({a_N\vec z})+\Omega(a_N\vec z) &\;\;{\rm if}\;\;0<\alpha\leq 1, \\
{\rm i} {a_N\vec z}\cdot\left<\vec V\right>+ I_{\alpha}({a_N\vec z})+\Omega(a_N\vec z) & \;\;{\rm if}\;\;1<\alpha\leq 2,\\
\displaystyle {\rm i} {a_N\vec z}\cdot\left<\vec V\right>+\frac{a_N^2}{2}\vec z \cdot\left(\mathbb{M}_{\vec V}\vec z\right)+\Omega(a_N\vec z) & \;\;{\rm if}\;\;2<\alpha\leq\infty.
\end{array}
\right.
\end{equation}
The renormalization conditions implies that
\begin{equation}
\label{new2}
\displaystyle \lim_{N\rightarrow\infty}a_N=0\;\;\Rightarrow\;\;{\lim_{a_N\rightarrow 0}}\frac{\Omega(a_N\vec z)}{I_\alpha(a_N\vec z)}=0\;\;(0<\alpha\leq 2),
\;\;\;\;{\lim_{a_N\rightarrow  0}}\frac{\Omega(a_N\vec z)}{a^2_N\vec z\cdot\left(\mathbb{M}_{\vec V}\vec z\right)}=0\;\;(2<\alpha<\infty),
\end{equation}
then the asymptotic expression for $\ln\left(\Psi_{\vec V}(a_N\vec z)\right)^N$ ($N\rightarrow\infty$) are given by: 
\begin{equation}
\label{new3}
N\ln\Psi_{\vec V}(a_N\vec z)\asymp
\left\{\begin{array}{cl}
{N I_{\alpha}(a_N\vec z)} & \;\;{\rm if}\;\;0<\alpha \leq 1, \\
{\rm i} {N a_N\vec z}\cdot\left<\vec V\right>+ N I_{\alpha}({a_N\vec z}) & \;\;{\rm if}\;\;1<\alpha\leq 2, \\
\displaystyle {\rm i} {N a_N\vec z}\cdot\left<\vec V\right>+\frac{N a_N^2}{2}\vec z \cdot\left(\mathbb{M}_{\vec V}\vec z\right) & \;\;{\rm if}\;\;2<\alpha\leq\infty.
\end{array}
\right.
\end{equation}
Finally, { we show} that
\begin{equation}
\label{new4}
\lim_{N\rightarrow\infty} N\ln\Psi_{\vec V}(a_N\vec z)=\lim_{N\rightarrow\infty}\psi(N,\vec z)\;\;\;\;{\forall {\vec z}\in\mathbb{R}^d}.
\end{equation}

The renormalization condition can be considered for all $N\geq N_0$, where $N_0$ is a given integer, or even asymptotically. For sake of simplicity, we consider that $N\geq 1$ for $\alpha\neq 1$ and $N>eK$ for $\alpha=1$.
Also, It is worth mentioning that for $0<\alpha\leq 2$ the functions $A_\alpha(\vec z)$ and $B_\alpha(\vec z)$ are obtained from the term $I_\alpha( \vec z)$ in the expansion of the characteristic function of  $\vec V$ given in (\ref{def4}). For $2<\alpha\leq \infty$ the matrix $\mathbb{M}_{\vec V}$ is the covariance matrix of the vector ${\vec V}$. {\it QED}

The scheme presented above to prove the theorem is simply Lévy's proof for random variables (see reference \cite{Lévy-1923}) generalized for random vectors.
For more details, see reference \cite{Figueiredo2019} and references therein. Other proofs can be found in the excellent book \cite{Meerschaert2001}.

We observe that the characteristic function $\psi_\alpha(N,{\vec z})$ represents a stable Lévy distribution for $0<\alpha\leq 2$ and a Gaussian one for $\alpha>2$. As expressed in the Theorem \ref{teo1}, it is directly related to the standard characteristic function of stable random vectors presented by Samorodnitsky and Taqqu in \cite{Taqqu}.

In the next theorem, we establish an asymptotic condition for a given { measure} $\rho_{\vec V}({\vec v})$ that allows to obtain the Lévy exponent $\alpha$, the function $A_\alpha(\hat{z})$ and the function $B_\alpha(\hat{z})$ given in Theorem \ref{teo1}.

\begin{teorema}
\label{teo2}
For a given direction ${\hat v}$, let us consider that the { measure} $\rho_{\vec V}({\vec v})$ satisfies the following asymptotic relationship  for $|\vec v|\rightarrow\infty$:
$$\rho_{\vec V}({\vec v})=\rho_{\vec V}({\hat v|\vec v|})\asymp\frac{C({\hat v})}{|{\vec v }|^\mu},\;\;\mu>d,\;\;{\forall {\hat v}\in S_d,}$$
where $C({\hat v})$ is  { an integrable function defined on $ S_d$}.
Then, the random vector ${\vec V}$ has Lévy exponent $\alpha=\mu-d$ and for $0<\alpha\leq 2$ the functions $A_\alpha(\hat z)$ and $B_\alpha(\hat z)$ are given by
$$A_{\alpha }\left(\hat{z}\right) =  \int _{S_{d} }C\left(\hat{v}\right)\left|\hat{z}\cdot \hat{v}\right|^{\alpha } d\Omega_{d}\, ;\;\;
B_{\alpha }\left(\hat{z}\right) = \int _{S_{d} }C\left(\hat{v}\right)
\frac{\hat{z}\cdot \hat{v}}{\left|\hat{z}\cdot \hat{v}\right|} \left|\hat{z}\cdot \hat{v}\right|^{\alpha } d\Omega_{d}.$$
\end{teorema}
\vspace*{2mm}
\noindent{\it Proof}: { Let us define the following sequence of measures for $n\in\mathbb{N}$:
$$
f_n({\vec v})=
\left\{
\begin{array}{c}
\rho_{\vec V}({\vec v})\;\;\;\;\;\;\mbox{if}\;\;\;|{\vec v}|\leq n, \\ \\
\displaystyle\epsilon_n\frac{C({\hat v})}{|{\vec v }|^\mu}\;\;\;\mbox{if}\;\;\;|{\vec v}|>n,
\end{array} 
\right.
\;\;\;\mbox{where}\;\;\; \int_{\mathbb{R}^d}f_n({\vec v})d^d{\vec v}=1.
$$
According to the reference \cite{Figueiredo2019}, the Lévy exponent of the measure $f_n({\vec v})$ is $\alpha_n=\alpha=d-\mu$ and its characteristic function $\psi_n({\vec z})$ has functions $I_{\alpha}^n({\vec z})$, which are introduced in definition 1, given by:
\begin{eqnarray}
I_{\alpha}^n \left(\vec{z}\right)&=&{\rm \; }-\frac{\pi }{\Gamma \left(\alpha +1\right)}
\frac{\cos \left({\alpha \pi  \mathord{\left/{\vphantom{\alpha \pi  2}}\right.\kern-\nulldelimiterspace} 2} \right)
A_{\alpha}^n \left(\hat{z}\right)-i\sin\left({\alpha \pi  \mathord{\left/{\vphantom{\alpha \pi  2}}\right.\kern-\nulldelimiterspace} 2} \right)
B_{\alpha}^n \left(\hat{z}\right)}{\sin\left(\alpha \pi \right)} \left|\vec{z}\right|^{\alpha }
\;\;(\forall \alpha \notin {\rm N}),\nonumber\\
I^n_{1}\left(\vec{z}\right) & = & \left[-\frac{\pi }{2} A_{1}^n \left(\hat{z}\right)+i\left[B_{1}^n \left(\hat{z}\right)
\left(1-\gamma -\ln u_c-\ln \left|\vec{z}\right|\right)-D_{1}^n \left(\hat{z}\right)\right]\right]\left|\vec{z}\right|,\nonumber\\
I^n_{2}\left(\vec{z}\right) & = & -\left[A^n_{2} \left(\hat{z}\right)\left(\frac{3}{2} -\gamma -\ln u_c-\ln \left|\vec{z}\right|\right)
-D^n_{2} \left(\hat{z}\right)+i\frac{\pi }{2} B^n_{2} \left(\hat{z}\right)\right]\frac{\left|\vec{z}\right|^{2}}{2},
\nonumber
\end{eqnarray}
where
\begin{eqnarray}
&A^n_{\alpha }\left(\hat{z}\right)  =  \epsilon_n\int _{S_{d} }C\left(\hat{u}\right)\left|\hat{z}\cdot \hat{u}\right|^{\alpha } dS_{d},\;\;\;
\displaystyle B^n_{\alpha}\left(\hat{z}\right)  = \epsilon_n\int _{S_{d} }C\left(\hat{u}\right)
\frac{\hat{z}\cdot \hat{u}}{\left|\hat{z}\cdot \hat{u}\right|} \left|\hat{z}\cdot \hat{u}\right|^{\alpha } dS_{d}, &
\nonumber\\
&D^n_{\alpha }\left(\hat{z}\right)  =  \epsilon_n\int _{S_{d} }C\left(\hat{u}\right)\left(\hat{z}\cdot \hat{u}\right)^{\alpha }
\ln\left|\hat{z}\cdot \hat{u}\right|dS_{d}.& \nonumber
\end{eqnarray}

From the definition of $f_n({\vec v})$ and the assumption made for the asymptotic behavior of $\rho_{\vec V}({\vec v})$, we have 
$$
\lim_{n\rightarrow\infty}f_n({\vec v})=\rho_{\vec V}({\vec v}),\;\;\;\lim_{n\rightarrow\infty}\psi_n({\vec z})=\psi_{\vec V}({\vec z}),
\;\;\;\lim_{n\rightarrow\infty}\epsilon_n=1.
$$  
Then, the functions $A_{\alpha}({\hat z})$ and $B_\alpha({\hat z})$ are obtained as
$$
A_{\alpha}({\hat z})=\lim_{n\rightarrow\infty}A^n_{\alpha }\left(\hat{z}\right),
\;\;\;B_\alpha({\hat z})=\lim_{n\rightarrow\infty}B^n_{\alpha}\left(\hat{z}\right).
$$
}
{\it QED}\\

\section{The problem of a test particle interacting with $N$ random particles}

Consider a test particle interacting with $N$ particles (called sources), where the force of each source on the test particle is central and inversely proportional to some positive power of the distance between them. Also,  let us consider a given system of units where $l_0$ denotes the length unit, $f_0$ is the force unit, and $k_0$ { the coupling} constant unit.
Thus, by considering the linear superposition of forces and explicitly stating system of units used, the expression of the force exerted on the test particle reads:
\begin{equation}
\label{forca_dimensao}
{\vec {\cal F}}^{N}_{ ren}= \sum_{i=1}^{N}{\vec {\cal F}}_{i}, \;\; 
{\vec{\cal F}}_{i}=k\,k_0\frac{{\vec y\,l_0}-{\vec y}_i\, l_0}
{\left|{\vec y}\,l_0-{\vec y}_i\,l_0\right|^{\delta+1}}=
k\frac{{\vec y}-{\vec y}_i}
{\left|{\vec y}-{\vec y}_i\right|^{\delta+1}}f_0,\;\;f_0= \frac{k_0}{l_0^\delta}\;(\delta>0),
\end{equation}
where ${\vec {\cal F}}_i$ ($i=1,\ldots,N$) is the force of the source $i$ on the test particle, $kk_0$ is the coupling constant, ${\vec y}\,l_0$ is the position vector of the test particle and ${\vec y}_i\,l_0$ the position vector of the source $i$. The position vectors ${\vec y},{\vec y}_i\in\mathbb{R}^d$ and the coupling constant $k\in\mathbb{R}$ are dimensionless. And we can define the dimensionless force as
\begin{equation}
\label{eq7}
{\vec f}_{ ren}^N(\vec y)=\frac{{\vec {\cal F}}^{N}_{ ren}}{f_0}= \sum_{i=1}^{N}{\vec f}_{i}(\vec y), \;\; 
{\vec f}_{i}(\vec y)=k\frac{{\vec y}-{\vec y}_i}
{\left|{\vec y}-{\vec y}_i\right|^{\delta+1}}.
\end{equation}
Henceforth in this work, we will only consider this dimensionless system of units.

The potential energy of the test particle due to its interaction with the sources can be defined as
\begin{equation}
\label{eq8}
u_{ ren}^{N}({\vec y}) = \sum_{i=1}^{N}{u}_{i}({\vec y}), \;\; 
u_i({\vec y})=\left\{
\begin{array}{cc}
\displaystyle \frac{k}{\delta-1}\frac{1}{\left|{\vec y}-{\vec y}_i\right|^{\delta-1}} & \mbox{if}\;\;\delta\neq 1, \\ \\
\displaystyle -k\ln \left|{\vec y}-{\vec y}_i\right| &  \mbox{if}\;\;\delta=1.
\end{array}\right.
\end{equation}
In order to obtain equation (\ref{eq8}) above, we simply use the definition of potential energy given by
\begin{equation}
\label{eq9}
{\vec f}_{ ren}^N=-\nabla_{\vec y}\left(u_{ ren}^N({\vec y})\right),\;\; {\vec f}_i({\vec y})=-\nabla_{\vec y}\left(u_i({\vec y})\right),
\end{equation}
where $\nabla_{\vec y}$ stands for the gradient operator with respect to the vector ${\vec y}$.

To model the randomness for the source positions, let us consider a random vector ${\vec Y}_i$ with { measure} $\rho_{{\vec Y}_i}\left( {\vec y}_i\right)$. Also, let us consider the position of the test particle ${\vec y}$ as being fixed. Hence, the force in (\ref{eq7}) can be considered as a random vector written as
\begin{equation}
\label{eq10}
{\vec F}^N_{ ren} = \sum_{i=1}^{N}k\frac{{\vec y}-{\vec Y}_i}
{|{\vec y}-{\vec Y}_i|^{\delta+1}}, 
\end{equation}
and the potential energy in (\ref{eq8}) as random variable given by
\begin{equation}
\label{eq11}
U_{ ren}^{N}
=\left\{
\begin{array}{cc}
\displaystyle\sum_{i=1}^{N} \frac{k}{\delta-1}\frac{1}{|{\vec y}-{\vec Y}_i|^{\delta-1}} & \mbox{if}\;\;\delta\neq 1, \\ \\
\displaystyle\sum_{i=1}^{N}-k\ln |{\vec y}-{\vec Y}_i| & \mbox{if}\;\;\delta=1.
\end{array}\right.
\end{equation}

For a given fixed position ${\vec y}$ of the test particle, let us consider the transformation ${\bf T}$  defined as ${\bf T}(\vec v)={\vec y}-{\vec v}$. Applying this transformation to the position vector of the sources and test particle, we  have respectively ${\vec x}_i={\bf T}({\vec y_i})={\vec y}-{\vec y}_i$ and ${\vec x}={\bf T}({\vec y})=\vec 0$.
Therefore, we define the random  vector ${\vec X}_i={\vec y}-{\vec Y}_i$ with { measure} $\rho_{{\vec X}_i}(\vec x_i)$, which represents the probability of the relative position of the origin with respect to the source $i$.
The expressions in eqs. (\ref{eq10}) and (\ref{eq11}) become
\begin{equation}
\label{eq12}
{\vec F}^N_{ ren} = \sum_{i=1}^{N}k\frac{{\vec  X}_i}
{|{\vec X}_i|^{\delta+1}}, \;\;
U_{ ren}^{N}
=\left\{
\begin{array}{cc}
\displaystyle\sum_{i=1}^{N} \frac{k}{\delta-1}\frac{1}{|{\vec X}_i|^{\delta-1}} & \mbox{if}\;\;\delta\neq 1, \\ \\
\displaystyle\sum_{i=1}^{N}-k\ln |{\vec X}_i| & \mbox{if}\;\;\delta=1.
\end{array}\right.
\end{equation}

Note that the gauge of the potential energy was chosen in such a way that it vanishes at $|{\vec x}_i|=\infty$ for $\delta>1$, at $|{\vec x_i}|= 1$ for $\delta=1$
and at $|{\vec x}_i|=0$ for $\delta<1$. Also, let us remark that the force on the test particle is  repulsive for $k>0$ and attractive for $k<0$.

Equation (\ref{eq12}) shows the force as a sum of $N$ random vectors and the  potential energy as a sum of $N$ random variables.
The problem of determining a well-defined limit for the probability distribution of these sums when $N\rightarrow\infty$ can be dealt with the results of the Central Limit Theorem presented in Theorem \ref{teo1}. However,
in order to apply these results,
we have to make some assumptions about the random vectors ${\vec X}_i$ and to define renormalization constants (which must depend on the number of particles $N$) with their respective renormalization vectors.

\begin{hipotese} {\rm ({\bf H1})}
{\rm (Independence and Identity)} The random vectors ${\vec X}_i$ are statistically independent and identical. Identity  means that the random vectors ${\vec X}_i$ are all equivalent to a same random vector denoted by ${\vec X}$. Indeed, this means that the probability distribution for the position of each source $i=1,\ldots,N$ are identical to the probability distribution of ${\vec X}$. Also, the source probability distribution $\rho_{\vec Y_i}({\vec y}_i)$ is  independent of the test particle position ${\vec y}$.
\end{hipotese}

\begin{definicao}
\label{def1}
{\rm (Renormalized Constants)} We define two renormalization constants.
\begin{enumerate}
\item The first is the coupling constant $k$. This means that the coupling constant may depend on the number of particles $N$ and to evidence this fact, we denote it as $k_N$.
\item The second is the size system $L_N$. Let us consider a given fixed random vector ${\vec R}$  with { measure} $\rho_{\vec R}({\vec r})$ in such a way that
${\bf X}_i \equiv L_N{\vec R}$.  
\end{enumerate}
\end{definicao}

It is worth to remark that the identity hypothesis made in H1 is fundamental to define just one renormalization size $L_N$, since from Definition \ref{def1}, we consider ${\vec X}_i\equiv L_N{\vec R}$.

\begin{definicao}
\label{def2}
{\rm (Force Renormalization Vector)}
Considering the random vector ${\vec R}$ in Definition \ref{def1}, then we define the force renormalization vector as ${\vec V}={\hat R}/|{\vec R}|^{\delta}$, where
${\hat R}={\vec R}/|{\vec R}|$.
\end{definicao}

\begin{definicao}
\label{def3}
{\rm (Energy Renormalization Variable)}
Considering the random variable $R=|{\vec R}|$ (with ${\vec R}$ given in Definition \ref{def1}), then we define the energy renormalization variable as
$$ U=\left\{\begin{array}{cc} R^{1-\delta} & {\rm if}\;\; \delta\neq 1, \\ \\ \displaystyle -\ln {R} & {\rm if}\;\; \delta=1.\end{array}\right.$$
The { measure}
$\rho_R(r)$ can be written as a function of the { measure} $\rho_{\vec R}({\vec r})$ as follows - see equation (\ref{eq5}) :
\begin{equation}
\label{eq13}
\rho_{R}(r)=r^{d-1}\int_{S_d}\rho_{\vec R}({\hat  r})d\Omega_d,\;\;{\hat r}=\frac{\vec r}{|\hat r|}.
\end{equation}
\end{definicao}

\begin{hipotese}
{\rm ({\bf H2})} The { measure} of the random vector ${\vec R}$ is  a continuous function at ${\vec r}=\vec 0$, that is,
$$\lim_{|{\vec r}|\rightarrow 0}\rho_{\vec R}({\vec r})=\rho_{\vec R}(\,\vec 0\,).$$
\end{hipotese}

\begin{hipotese}
{\rm ({\bf  H3})} For $\delta \leq 1$, we assume that
\begin{eqnarray}
&&\int_{\mathbb{R}^d}|{\vec r}|^{2(1-\delta)}\rho_{\vec R}({\vec r})d^d{\vec r }<\infty\;\;\;\;{\rm if}\;\;\;\;\delta<1,\nonumber\\
&&\int_{\mathbb{R}^d}\left(\ln|{\vec r}|\right)^2\rho_{\vec R}({\vec r})d^d{\vec r}<\infty\;\;\;\;{\rm if}\;\;\;\;\delta=1.\nonumber
\end{eqnarray}
\end{hipotese}

Considering H1 and the transformation ${\vec X}_i={\vec y}-{\vec Y}_i$, we can see that the identity and independence between the random vectors ${\vec X}_i$ stem from the identity and independence between the vectors ${\vec Y}_i$. On the other hand, the independence of ${\vec Y}_i$, with respect to the test particle position, implies that ${\vec X}_i$ depend on the position of the test particle through the linear transformation ${\vec X_i}={\vec y}-{\vec Y}_i$. The relationship between their respective characteristic functions is given by:
\begin{equation}
\Psi_{{\vec X}_i}({\vec z})=\left<e^{{\vec X}_i\cdot{\vec z}}\right>=\left<e^{({\vec y} -{\vec Y}_i)\cdot{\vec z}}\right>=
e^{{\vec y}\cdot\vec z}\left<e^{-{\vec Y}_i\cdot{\vec z}}\right>=e^{{\vec y}\cdot\vec z}\Psi_{{\vec Y}_i}(-\vec z).
\end{equation}

Considering H2 and H3, there are no assumptions about the global isotropy or uniformity of the source probability distribution. On the one hand, H2 and the transformation ${\vec X}_i={\vec y}-{\vec Y}_i$ just imply that the { measure} $\rho_{{\vec Y}_i}({\vec y }_i)$ is a continuous function at ${\vec y}_i={\vec y}$. It means that for a sufficiently small neighborhood of the test particle, the { measure} is approximately constant. Mathematically, we can interpret this property as local isotropy and uniformity for the source distribution around the test particle. On the other hand, H3 only imposes the finitude for the variance of $|\vec R|$ without providing conditions about its global isotropy or uniformity.

With these definitions and assumptions, it is quite straightforward to show that the force in (\ref{eq12}) can be written as the following sum of independent and identical random vectors:
\begin{equation}
\label{eq14}
\vec{F}_{ ren}^{N} = \sum_{i=1}^{N}\vec{F}^{N}_i, \;\;
\vec{F}^{N}_i\equiv a_N{\vec V},\;\;
 a_{N}=\frac{k_N}{L_{N}^\delta}.
\end{equation}
Therefore, its respective characteristic function can be given by:
\begin{equation}
\label{eq15}
\Psi_{\vec{F}_{ ren}^{N}}(\vec{z}) = \prod_{i=1}^{N} \Psi_{\vec{F}_{i}^{N}}(\vec{z}) =\Psi_{\vec{V}}(a_{N}\vec{z})^{N},
\end{equation}
where we have used the convolution property in eq. (\ref{independencia}) for the sum of independent random vectors.

Similarly, the  potential energy (\ref{eq12}) is  expressed as follows:
\begin{equation}
\label{eq16}
U_{ ren}^N=\sum_{i=1}^N U^N_i,\;\;
U_i^N\equiv\left\{
\begin{array}{ll}
b_NU\;\;\left(\displaystyle b_N=\frac{a_NL_N}{\delta-1}\right) & \hspace*{-20mm}\mbox{if}\;\;\delta\neq 1, \\ \\
 -b_N\ln L_N+b_NU\;\;(b_N=a_NL_N) & \mbox{if}\;\;\delta=1,
\end{array}
\right.
\end{equation}
with the respective characteristic function given by:
\begin{equation}
\label{eq17}
\Psi_{U_{ ren}^{N}}(z) = \prod_{i=1}^{N} \Psi_{U_{i}^{N}}(z) =
\left\{\begin{array}{cc}
\Psi_{U}(b_{N}{z})^{N} & \mbox{if}\;\;\delta\neq 1, \\ \\
e^{-INb_N\ln L_N}\Psi_{U}(b_{N}{z})^{N} & \mbox{if}\;\;\delta=1.
\end{array}
\right.
\end{equation}

\section{The Limits for the Renormalized  Force and Potential Energy}

The purpose of this section is to apply the Theorems 1 and 2 to state two new theorems about the limit $N\rightarrow\infty$ for the force (Theorem \ref{teo3})  and potential energy (Theorem \ref{teo4}) given in equations (\ref{eq14}$-$\ref{eq17}).

\subsection{The Limit for the Renormalized  Force}

In order to formulate the Theorem \ref{teo3} in terms of random vectors instead of characteristic functions, let us denote ${\vec S}_{\alpha}$ as being the symmetric $\alpha$-stable vector
with characteristic function given by:
\begin{equation}
\label{Lévy_simetrica}
\ln\Psi_{{\vec S}_\alpha}(\vec z)=-\frac{|\vec z|^\alpha}{\Gamma(\alpha+1)},\;\; 0<\alpha\leq2.
\end{equation} 
Also, for a given random vector ${\vec V}$ with finite covariance matrix let us denote ${\vec G}_{\vec V}$ as being the random vector with Gaussian characteristic function:
\begin{equation}
\label{gaussiana}
\ln\Psi_{{\vec G}_{\vec V}}({\vec z})=-\frac{1}{2}\vec z\cdot\left(\mathbb{M}_{\vec V}\vec z\right),
\end{equation}
where $\mathbb{M}_{\vec V}$ is the covariance matrix of ${\vec V}$.

\begin{teorema} 
\label{teo3}
Let us consider the characteristic function limit for the renormalized  force vector given in equation (\ref{eq15}):
$$\phi_{{\vec F}_{ ren}}(\vec z)=\lim_{N\rightarrow\infty}\Psi_{\vec{F}_{ ren}^{N}}(\vec{z}) = \lim_{N\rightarrow\infty}\Psi_{\vec{V}}(a_{N}\vec{z})^{N}.$$
Let us assume the hypothesis H1 and H2, then the asymptotic expression for the  force when $N\rightarrow\infty$ is given by
\begin{equation}
\label{eq22}
{\vec F}^N_{ ren}\asymp
\left\{
\begin{array}{ccll}
\displaystyle \left(\frac{\lambda_{\alpha}\rho_{\vec R}(\,\vec 0\,)K}{\delta}\right)^{1/\alpha}{\vec S}_\alpha &&  & {\rm if}\;\; 0<\alpha<1\;(d<\delta),
\\ \\
\displaystyle \sigma_N \frac{\lambda_1\rho_{\vec R}(\,\vec 0\,)}{\delta}{\vec S}_1 &&  &  {\rm if}\;\;\alpha=1\;(\delta=d),
\\ \\
\displaystyle \langle qK\vec V\rangle-\sigma_N
\left(\frac{\lambda_{\alpha}\rho_{\vec R}(\,\vec 0\,)}{\delta}\right)^{1/\alpha}{\vec S}_\alpha &&   &  {\rm if}\;\;1<\alpha\leq 2\;(\delta<d<2\delta),
\\ \\
\displaystyle \langle qK\vec V\rangle+\sigma_N{\vec G}_{\vec V}
 &&   &  {\rm if}\;\;2<\alpha\leq \infty\;(2\delta<d).
\end{array}\right. 
\end{equation}
Moreover, the limit for ${\vec{F}_{ ren}^{N}}$  is given by
\begin{equation}
\label{eq22a}
\lim_{N\rightarrow\infty}{\vec F}^N_{ ren}={\vec F}_{ ren}=
\left\{
\begin{array}{ccll}
\displaystyle \left(\frac{\lambda_{\alpha}\rho_{\vec R}(\,\vec 0\,)K}{\delta}\right)^{1/\alpha}{\vec S}_\alpha &&  &  {\rm if}\;\;0<\alpha<1\;(d<\delta),
\\ \\
\vec 0 &&  &  {\rm if}\;\;\alpha=1\;(\delta=d),
\\ \\
\displaystyle \langle qK\vec V\rangle &&   &  {\rm if}\;\;1<\alpha\leq \infty\;(\delta<d),
\end{array}\right. 
\end{equation}
where $q=a_N/|a_N|$, $\lambda_\alpha$ is a positive number depending on the Lévy exponent given by $\alpha=d/\delta$. The values for $K>0$ and $\sigma_N$ for the different {ranges} of $\alpha$  are given in Table \ref{tabela3}.
\begin{table}[h]
\begin{center}
\begin{tabular}{ccccc|ccccc|ccccc}
\hline\hline
& $0<\alpha<1$ & & & & & & $\alpha=1$  & & & & & &  $1<\alpha<\infty$  &\\ 
& $\displaystyle N|a_N|^\alpha=K$ & & & & & &  $\displaystyle - N|a_N|\ln |a_N|=K$ & & & & & & 
$\displaystyle N|a_N|=K$ & \\
\hline\hline
\end{tabular}
\begin{tabular}{c}
\begin{tabular}{ccc|ccc|cccc|ccc}
\\ \hline\hline
$\alpha=1$ & & & & $1<\alpha<2$ & & & & $\alpha=2$ & & & & $2<\alpha\leq\infty$\\
$\sigma_N=Nh(K/N)$ & & & & $\sigma_N=KN^{(1-\alpha)/\alpha}$ & & & & $\sigma_N=K\left(N^{-1}\ln N\right)^{1/2}$ & & & &  $\displaystyle\sigma_N=\frac{KN^{-1/2}}{2}$ \\
\hline\hline
\end{tabular}
\end{tabular}
\end{center}
\vspace*{-5mm}
\caption{The values for $K>0$ and $\sigma_N$ for the different {ranges} of $\alpha$.}
\label{tabela3}
\end{table}

\end{teorema}

\noindent{\it Proof:} The renormalization vector ${\vec V}$ in Definition \ref{def2} is given by an invertible differentiable transformation of  ${\vec R}$, then  equation ({\ref{eq2}) and hypothesis H2 implies the following asymptotic relation for ${\vec r}\rightarrow \vec 0$:
\begin{equation}
\label{eq19}
\rho_{\vec{V}}(\vec{v})\asymp \rho_{\vec{R}} (\vec{r})\left\|\frac{\partial\vec{r}}{\partial\vec{v}}\right\|\asymp
\frac{1}{\delta}\,\frac{\rho_{\vec{R}}(\hat v/|\vec{v}|^{1/\delta})}{|\vec v|^{d+d/\delta}},\;\;{\rm where}\;\;\vec{v}=\frac{\hat{r}}{|\vec{r}|^\delta}.
\end{equation}
Then, for $|{\vec v}|\rightarrow\infty$, we obtain:
\begin{equation}
\label{eq20}
\rho_{\vec{V}}(\vec{v})\asymp\frac{1}{\delta|\vec v|^{d+d/\delta}}\lim_{|\vec r|\to 0}\rho_{\vec{R}}(\vec{r})\asymp\frac{\rho_{\vec R}(\,\vec 0\,)}{\delta|\vec v|^{d+d/\delta}}.
\end{equation}
From Theorem \ref{teo2}, we obtain the Lévy exponent $\alpha$ and the function $C({\hat v})$:
\begin{equation}
\label{eq21}
\alpha=\frac{d}{\delta},\;\;C({\hat v})=\frac{\rho_{\vec R}(\,\vec 0\,)}{\delta}.
\end{equation}
Using the expression for $C(\hat v)$ given (\ref{eq21}), we obtain the functions $A_\alpha(\hat z)$, $B_\alpha(\hat z)$ defined in Theorem \ref{teo2}:
\begin{equation}
\label{teo3_1}
A_{\alpha }\left(\hat{z}\right) = \frac{\rho_{\vec R}(\,\vec 0\,)}{\delta} \int _{S_{d} }\left|\hat{z}\cdot \hat{v}\right|^{\alpha } d\Omega_{d}\, \;\;
B_{\alpha }\left(\hat{z}\right) =\frac{\rho_{\vec R}(\,\vec 0\,)}{\delta}\int _{S_{d} }
\frac{\hat{z}\cdot \hat{v}}{\left|\hat{z}\cdot \hat{v}\right|} \left|\hat{z}\cdot \hat{v}\right|^{\alpha } d\Omega_{d}=0.
\end{equation}
Therefore, the function $G_\alpha(\hat z)$ defined in Theorem \ref{teo1} can be written as
\begin{equation}
\label{teo3_2}
 G_\alpha(\hat z)=\frac{\rho_{\vec R}(\,\vec 0\,)}{\delta} \pi\frac{\cos(\alpha\pi/2)}{\sin(\alpha\pi)}\int _{S_{d} }\left|\hat{z}\cdot \hat{v}\right|^{\alpha } d\Omega_{d}.
\end{equation}
Considering the functions in eq. (\ref{teo3_1}) and (\ref{teo3_2}) in Theorem \ref{teo1}, we can write the following asymptotic expression for the characteristic function $\Psi_{\vec V}(a_N\vec z)^N$
when $N\rightarrow\infty$:
\begin{equation}
\label{eq18teo3a}
\ln\psi_{\vec V}({a_N\vec z})^N\asymp
\left\{\hspace*{2mm}
\begin{array}{ll}
\hspace*{-2mm}\displaystyle -\frac{|\vec{z}|^\alpha}{\Gamma(\alpha +1)}\frac{\rho_{\vec R}(\,\vec 0\,)K}{\delta} \pi\frac{\cos(\alpha\pi/2)}{\sin(\alpha\pi)}\int _{S_{d} }\left|\hat{z}\cdot \hat{v}\right|^{\alpha } d\Omega_{d} &  {\rm if}\;\;0<\alpha<1,
\\ \\
\hspace*{-2mm}\displaystyle- \frac{|\vec z|}{\Gamma(\alpha+1)} \sigma_N\frac{\rho_{\vec R}(\,\vec 0\,)}{\delta}\frac{\pi}{2} \int _{S_{d} }\left|\hat{z}\cdot \hat{v}\right| d\Omega_{d} &  {\rm if}\;\;\alpha=1,
\\ \\
\hspace*{-2mm}\displaystyle{\rm i}\vec z\cdot\langle qK\vec V\rangle-\frac {|\vec{z}|^\alpha }{\Gamma(\alpha +1)}\sigma_N^\alpha\frac{\rho_{\vec R}(\,\vec 0\,)}{\delta} \pi\frac{\cos(\alpha\pi/2)}{\sin(\alpha\pi)}\int _{S_{d} }\left|\hat{z}\cdot \hat{v}\right|^{\alpha } d\Omega_{d} &  {\rm if}\;\;1<\alpha< 2,
\\ \\
\hspace*{-2mm}\displaystyle{\rm i}\vec z\cdot\langle qK\vec V\rangle-\frac {|\vec{z}|^2 }{\Gamma(\alpha+1)}\sigma_N^2\frac{\rho_{\vec R}(\,\vec 0\,)}{\delta}\frac{1}{2}\int _{S_{d} }\left|\hat{z}\cdot \hat{v}\right|^{2} d\Omega_{d} &  {\rm if}\;\;\alpha= 2,
\\ \\
\hspace*{-2mm}\displaystyle{\rm i}\vec z\cdot \langle{qK\vec V}\rangle-\frac{\sigma_N^2}{2}\vec{z}\cdot\left(\mathbb{M}_{\vec V}\vec{z}\right) &  {\rm if}\;\;2<\alpha<\infty) .
\end{array}\right.
\end{equation} 
Hence, tacking into account that $\Psi_{{\vec F}^N_{ ren}}=\psi_{\vec V}({a_N\vec z})^N$, we have for $N\rightarrow\infty$:
\begin{equation}
\label{eq18teo3b}
\ln\Psi_{{\vec F}^N_{ ren}}(\vec z)\asymp
\left\{\hspace*{2mm}
\begin{array}{ll}
\hspace*{-2mm}\displaystyle -\frac{\lambda_\alpha\rho_{\vec R}(\,\vec 0\,)K}{\delta}\frac{|\vec{z}|^\alpha}{\Gamma(\alpha +1)}  &  {\rm if}\;\;0<\alpha<1,
\\ \\
\hspace*{-2mm}\displaystyle- \sigma_N\frac{\lambda_\alpha\rho_{\vec R}(\,\vec 0\,)}{\delta}\frac{|\vec z|}{\Gamma(\alpha+1)}  &  {\rm if}\;\;\alpha=1,
\\ \\
\hspace*{-2mm}\displaystyle{\rm i}\vec z\cdot\langle qK\vec V\rangle-\sigma_N^\alpha\frac{\lambda_\alpha\rho_{\vec R}(\,\vec 0\,)}{\delta}\frac {|\vec{z}|^\alpha }{\Gamma(\alpha +1)}  & {\rm if}\;\;1<\alpha\leq 2,
\\ \\
\hspace*{-2mm}\displaystyle{\rm i}\vec z\cdot \langle{qK\vec V}\rangle-\frac{\sigma_N^2}{2}\vec{z}\cdot\left(\mathbb{M}_{\vec V}\vec{z}\right) &  {\rm if}\;\;2<\alpha<\infty),
\end{array}\right.
\end{equation} 
where we define:
\begin{equation}
\label{new5}
\lambda_\alpha= \pi\frac{\cos(\alpha\pi/2)}{\sin(\alpha\pi)}\int _{S_{d} }\left|\hat{z}\cdot \hat{v}\right|^{\alpha } d\Omega_{d}\;\; {\rm if}\;\;0<\alpha< 2),\;\;\;\;
\lambda_2=\frac{1}{2}\int _{S_{d} }\left|\hat{z}\cdot \hat{v}\right|^{2} d\Omega_{d}.
\end{equation}
Finally, considering the property in eq. (\ref{fc_renormalizacao}), the characteristic functions in eqs. (\ref{Lévy_simetrica}) and (\ref{gaussiana}), we can show the asymptotic relations exhibited in (\ref{eq22}). Considering that $\lim_{N\rightarrow\infty}\sigma_N=0$ and tacking the limit $N\rightarrow\infty$ in (\ref{eq18teo3b}), we obtain (\ref{eq22a}).
{\it QED}

The analysis of the results presented in the Theorem \ref{teo3} shows that for the interval $0<\delta<d$ ($\alpha>1$), the force converges to its mean when the number of sources grows to the infinite.
Moreover, for very large $N$, the force is asymptotically characterized by an average value added to { an} $\alpha$-stable Lévy vector for $1<\alpha\leq 2$ and a Gaussian vector for $\alpha >2$. We can see that these vectors characterize the stochastic fluctuations of the force around its average value. These fluctuations have a width modulated by the parameter $\sigma_N$ then this width decreases to zero as the number of sources grows to infinite.

Looking at this problem in terms of the force { measure}, we see that it converges to a $\delta$-Dirac { measure} centered on the average force. For each value of $\alpha>1$, we have a sequence of different $\alpha$-stable or Gaussian measures that converge to a $\delta$-Dirac measure.
As the force field on the test particle converges to its mean value, we can call it the Mean Field Limit. This type of limit is analogous to the Vlasov limit in the study of a system of $N$  interacting particles.

For the interval $\delta>d$ ($\alpha<1$), the force converges to { an} $\alpha$-stable vector with an infinite mean. Here, the force fluctuates according to a purely stochastic quantity, making it impossible to be approximated by an average value. Known as the Stochastic Limit, it emerges as the opposite of the Mean Field Limit. Be aware that if the force does not converge, its randomness is an irremovable property, even considering that the number of random sources increases indefinitely. In this case, no mean field theory can explain the nature of the force.

The singular situation $\delta=d$ ($\alpha=1$) represents an intermediate case between the other two limits previously considered. Although the force does not have a well-defined average for each value of $N$, the force fluctuates stochastically around zero with smaller and smaller amplitude as $N\rightarrow\infty$. We have an asymptotic behavior similar to what we call Mean Field Limit, where the force asymptotically converges to a well-defined value but which, in this singular case, is not determined by the mean of the force.

\subsection{The Limit for the  Potential Energy}

The potential energy is a real random variable (instead of a random vector, as is the force). Then, to apply the Theorem \ref{teo1} to calculate the potential energy, we must identify real random variables as one-dimensional random vectors. To express the results of Theorem \ref{teo4} below directly in terms of random variables and not in terms of their respective characteristic functions, let us define some special stable random variables.

We denote $S_{\alpha'}^*$  for $0<\alpha'<2$ ($\alpha'\neq 1,2$) as being the random variable with asymmetric $\alpha'$-stable distribution whose characteristic function is given by 
\begin{equation}
\label{Lévy_assimetrica}
\ln\Psi_{S^*_{\alpha'}}(z)=-\frac{|z|^{\alpha'}}{\Gamma(\alpha' +1)}\left(1-i\beta\frac{z}{|z|}\tan\left(\frac{\alpha'\pi}{2}\right)\right),
\end{equation}
where $\beta$, called asymmetry parameter, may assume two values: $-1$ or $1$.

We denote by $S_{\alpha'}$ for $0<\alpha'\leq 2$ the random variable that has the following characteristic function associated with a $\alpha'$-stable symmetric distribution:
\begin{equation}
\label{Lévy_estavel1d}
 \ln\Psi_{S_{\alpha'}}(z)=-\frac{|z|^{\alpha'}}{\Gamma(\alpha' +1)}.
\end{equation}

Finally, for a random variable $U$ with finite variance $\left<U^2\right>=\sigma_U^2<\infty$, we denote $G_U$ as being the Gaussian random variable with characteristic function given by
\begin{equation}
\label{gaussiana1d}
\ln\Psi_{G_U}(z)=-\frac{\sigma_U z^2}{2}.
\end{equation}

Taking into account these random variables defined above, we can formulate the following theorem.
\begin{teorema}
\label{teo4}
Let us consider the characteristic function limit for the  potential energy in equation (\ref{eq17}):
$$
\lim_{N\rightarrow\infty}\Psi_{U_{ ren}^{N}}(z) =
\left\{\begin{array}{cc}
\displaystyle \lim_{N\rightarrow\infty}\Psi_{U}(b_{N}{z})^{N} & \mbox{if}\;\;\delta\neq 1, \\ \\
\displaystyle \lim_{N\rightarrow\infty}e^{-{\rm i}Nb_N\ln L_N}\Psi_{U}(b_{N}{z})^{N} & \mbox{if}\;\;\delta=1,
\end{array}
\right.
$$
Let us assume the hypothesis H1,  H2 for $\delta>1$ and H3 for $\delta\leq 1$, then the asymptotic expression for the potential when $N\rightarrow\infty$ is given by
\begin{equation}
\label{eq34}
U_{ ren}^N\asymp
\left\{
\begin{array}{lcl}
\displaystyle \left(\frac{\lambda'_{\alpha'}\Omega_d\rho_{\vec R}(\,\vec 0\,)K'}{\delta-1}\right)^{1/\alpha'}S_{\alpha'}^* &  &  {\rm if}\;\;0<\alpha'<1\;(\delta>d+1),
\\ \\
\displaystyle {\rm i} z\left(-\frac{\Omega_d\rho_{\vec R}(\,\vec 0\,)K'}{\delta-1}\right)+\sigma'_N\left(\frac{\lambda'_{1}\Omega_d\rho_{\vec R}(\,\vec 0\,)K'}{\delta-1}\right)S_{\alpha'}&  &  {\rm if}\;\;\alpha'=1\;(\delta=d+1),
\\ \\
\displaystyle {\rm i}{z}\langle q'K' U\rangle+\displaystyle \sigma'_N\left(\frac{\lambda'_{\alpha'}\Omega_d\rho_{\vec R}(\,\vec 0\,)K'}{\delta-1}\right)^{1/\alpha'}S_{\alpha'}^*&  &  {\rm if}\;\;1<\alpha'<2\;
\displaystyle \left(\frac{d}{2}+1<\delta<d+1\right),
\\ \\
\displaystyle {\rm i}{z}\langle q'K' U\rangle+\displaystyle \sigma'_N\left(\frac{\lambda'_{2}\Omega_d\rho_{\vec R}(\,\vec 0\,)K'}{\delta-1}\right)^{1/2}S_{2}&  &  {\rm if}\;\;\alpha'=2\;
\displaystyle\left(\delta=\frac{d}{2}+1\right),
\\ \\
\displaystyle {\rm i}{z}\langle q'K' U\rangle+\sigma'_N G_U & &  {\rm if}\;\;\alpha'>2\;\displaystyle\left(\delta<\frac{d}{2}+1\right)\; (\delta\neq 1),
\\ \\
\displaystyle {\rm i}{z}\left(-q'K'\ln L_N+\langle q'K' U\rangle\right)+\sigma'_N G_U & &  {\rm if}\;\;\alpha'>2\; (\delta=1).
\end{array}\right.\, 
\end{equation}
Moreover, the limit for ${\vec{U}_{ ren}^{N}}$  is given by
\begin{equation}
\label{eq34b}
\lim_{N\rightarrow\infty}U_{ ren}^N=U_{ ren}=
\left\{
\begin{array}{lcl}
\displaystyle \left(\frac{\lambda_{\alpha'}\Omega_d\rho_{\vec R}(\,\vec 0\,)K'}{\delta-1}\right)^{1/\alpha'}S_{\alpha'}^* &  &  {\rm if}\;\;0<\alpha'<1\;(\delta>d+1),
\\ \\
\displaystyle {\rm i} z\left(-\frac{\Omega_d\rho_{\vec R}(\,\vec 0\,)K'}{\delta-1}\right)&  &  {\rm if}\;\;\alpha'=1\;(\delta=d+1),
\\ \\
\displaystyle {\rm i}{z}\langle q'K' U\rangle & &  {\rm if}\;\;\alpha'>1\; (\delta<d+1)\; (\delta\neq 1),
\\ \\
\displaystyle {\rm i}{z}\left(-q'K'\lim_{N\rightarrow\infty}\ln L_N+\langle q'K' U\rangle\right) & &  {\rm if}\;\;\alpha'>1\;(\delta<d+1)\; (\delta=1),
\end{array}\right.\, 
\end{equation}
where $q'=b_N/|b_N|$, $\lambda'_{\alpha'}$ is a positive number depending on the Lévy exponent given by $\alpha'=d/(\delta-1)$ for $\delta>1$. For $\delta\leq 1$, we have $\alpha'> 2$. The values for $K'>0$ and $\sigma'_N$ for the different {ranges} of $\alpha'$  are given in the Table \ref{tabela4}.
\begin{table}[h]
\begin{center}
\begin{tabular}{ccccc|ccccc|ccccc}
\hline\hline
& $0<\alpha'<1$ & & & & & & $\alpha'=1$  & & & & & &  $1<\alpha'<\infty$  &\\ 
& $\displaystyle N|b_N|^\alpha=K'$ & & & & & &  $\displaystyle -N|b_N|\ln |b_N|=K'$ & & & & & & 
$\displaystyle N|b_N|=K'$ & \\
\hline\hline
\end{tabular}
\vspace*{0mm}
\begin{tabular}{c}
\begin{tabular}{ccc|ccc|cccc|ccc}
\\ \hline\hline
$\alpha'=1$ & & & & $1<\alpha'<2$ & & & & $\alpha'=2$ & & & & $2<\alpha'\leq\infty$\\
$\sigma'_N=Nh(K'/N)$ & & & & $\sigma'_N=K'N^{(1-\alpha')/\alpha'}$ & & & & $\sigma'_N=K'\left(N^{-1}\ln N\right)^{1/2}$ & & & &  $\displaystyle\sigma'_N=\frac{K'N^{-1/2}}{2}$ \\
\hline\hline
\end{tabular}
\end{tabular}
\end{center}
\vspace{-5mm}
\caption{The values for $K'>0$ and $\sigma'_N$ for the different {ranges} of $\alpha'$.}
\label{tabela4}
\end{table}
\end{teorema}

\noindent {\it Proof:}
We should apply Theorems \ref{teo1} and \ref{teo2} to calculate the limit for $\psi_{U}(b_Nz)^N$ when $N\rightarrow\infty$, where $U$ is the energy renormalization variable in Definition \ref{def3}.
Let us define a random variable ${\bar U}$ as being 
\begin{equation}
\label{eq26}
{\bar U}=
\left\{
\begin{array}{cc}
U & \;\; {\rm if}\;\;{\bar U}\geq 0, \\
0 & \;\; {\rm if}\;\;{\bar U}<0.
\end{array}\right.\;\;\;\Leftrightarrow\;\;\; 
\rho_{\bar U}(\bar u)=
\left\{
\begin{array}{cl}
\rho_{U}(u) & \;\; {\rm if}\;\;{\bar u}=u\geq 0, \\
0 & \;\; {\rm if}\;\; {\bar u}<0.
\end{array}\right.\, 
\end{equation}
Therefore, the respective probability distributions for $\bar U$ and $U$ are identical since they have the same characteristic function. We define $\bar U$ only as technical support to apply Theorems \ref{teo1} and \ref{teo2}. In fact, according to Definition \ref{def3}, the random variable $U$ cannot assume negative values, which is equivalent to defining the variable ${\bar U}$ as done in (\ref{eq26})
\\

\noindent The obtaining of the Lévy exponent $\alpha'$ and the respective characteristic function splits in two cases: (i) $\delta>1$ and (ii) $\delta\leq 1$.

(i) $\delta>1$ $-$ Taking into account ${\bar U}>0$ in (\ref{eq26}),  hypothesis H2 and equation (\ref{eq1}), we have asymptotically for $r\rightarrow 0$ (respectively $u\rightarrow\infty)$:
\begin{equation}
\label{eq27}
 \rho_{\bar U}(\bar u)=\rho_{U}(u)\asymp \rho_R(r)\left|\left|\frac{\partial r}{\partial u}\right|\right|\asymp \frac{1}{\delta-1}\rho_R(r)r^{\delta}\asymp \frac{1}{\delta-1}\rho_R\left( u^{1/(1-\delta)} \right)u^{\delta/(1-\delta)}.
\end{equation}
Therefore, considering the asymptotic expression for  $u\rightarrow\infty$ in (\ref{eq27}),  we get
\begin{equation}
\label{eq28}
\rho_{\bar U}(\bar u)\asymp\rho_U(u)\asymp
\displaystyle \frac{1}{\delta-1}\left[ \lim_{u\rightarrow \infty}\rho_R\left( u^{1/(1-\delta)}\right) \right]u^ {\delta/(1-\delta)}.
\end{equation}
Using the expression for $\rho_R(r)$ given  in (\ref{eq13}), we obtain
\begin{equation}
\label{eq29}
 \rho_R\left( u^{1/(1-\delta)}\right)\asymp\rho_R(r)\asymp\left[r^{d-1}\int_{S_d}
\rho_{\vec R}({\hat r})d\Omega_d\right]=\Omega_d\rho_{\vec R}(\,\vec 0\,)u^{(d-1)/(1-\delta)},
\end{equation}
where $\Omega_d$ is given in (\ref{eq6}). 
Substituting (\ref{eq29}) into (\ref{eq28}) and considering equation (\ref{eq26}) for ${\bar u}<0$, we obtain:
\begin{equation}
\label{eq30}
\rho_{\bar U}(\bar u)\asymp\left\{
\begin{array}{cl}
\displaystyle \frac{\Omega_d\rho_{\vec R}(\,\vec 0\,)}{\delta-1}\frac{1}{{\bar u}^{1+d/(\delta-1)}} & \;\;{\rm if}\;\;\bar u>0\;(\bar u = u) \\
0 & \;\;{\rm if}\;\;\bar u<0.
\end{array}\right.\; 
\end{equation}
In order to apply Theorem \ref{teo2}, we rewrite the expression (\ref{eq30}) as follows:
\begin{equation}
\label{eq31}
\rho_{\bar U}({\bar u})\asymp\frac{C({\hat u})}{|\bar u|^{1+d/(\delta-1)}},\; \;\
C({\hat u})=
\left\{
\begin{array}{cl}
\displaystyle \frac{\Omega_d\rho_{\vec R}(\,\vec 0\,)}{\delta-1} & \;\;{\rm if}\;\;{\hat u}=1,\\
0 & \;\;{\rm if}\;\;\hat{u}=-1.
\end{array}\right.\, 
\end{equation}
Therefore, we can conclude that  $\alpha'=d/(\delta-1)$ for $\delta>1$.
In particular, for $0<\alpha'\leq 2$, we can obtain
\begin{eqnarray}
\label{eq35}
A_{\alpha'}\left(\hat{z}\right)&=&\int _{S_{d} }C\left(\hat{u}\right)\left|\hat{z}\cdot \hat{u}\right|^{\alpha'}d\Omega_d=C(1)=\frac{\Omega_d\rho_{\vec R}(\,\vec 0\,)}{\delta-1}, \nonumber\\
B_{\alpha'}\left(\hat{z}\right)&=& \int _{S_{d} }C\left(\hat{u}\right)
\frac{\hat{z}\cdot \hat{u}}{\left|\hat{z}\cdot \hat{u}\right|} \left|\hat{z}\cdot \hat{u}\right|^{\alpha'}d\Omega_d=q'C(1)=\frac{\Omega_d\rho_{\vec R}(\,\vec 0\,)}{\delta-1}\frac{z}{|z|},
\end{eqnarray}
where we have used the expression for $C({\hat u})$ given in (\ref{eq31}) into the expressions for $A_{\alpha'}(\hat z)$ and $B_{\alpha'}(\hat z)$ given in Theorem \ref{teo2}.
Plugging the functions given in (\ref{eq35}) into the function $G_{\alpha'}(\hat z)$  defined in Theorem \ref{teo1} for $0<\alpha'<2$ ($\alpha'\neq 1$), we obtain:
\begin{equation}
\label{eq36}
G_{\alpha'}({\hat z})=\frac{\lambda'_{\alpha'}\Omega_d\rho_{\vec R}(\,\vec 0\,)}{\delta-1}\left(1-iq'\frac{z}{|z|}\tan\left(\frac{\alpha'\pi}{2}\right)\right).\;\;
\lambda'_{\alpha'}=\pi\frac{\cos(\alpha'\pi/2)}{\sin(\alpha'\pi)}.
\end{equation}

(ii) $\delta\leq 1$ $-$  From the definitions for $U$, ${\bar U}$ and  $\rho_R(r)$ given in (\ref{eq13}), we have
\begin{eqnarray}
\label{eq32}
&\displaystyle \langle U^2\rangle=\int_{-\infty}^{\infty}\bar u^2\rho_{\bar U}(\bar u)d\bar u=\int_{0}^{\infty}u^2\rho_{U}(u)du=\int_{0}^{\infty}u^2(r)\rho_{R}( r)dr=\int_{0}^{\infty} u^2(r)r^{d-1}dr\int_{S_d}\rho_{\vec R}({\hat r})d\Omega_d,& \nonumber \\
& u(r)=\left\{\begin{array}{cc} r^{1-\delta} & \;\;{\rm if}\;\;\delta<1,\\ -\ln r & \;\;{\rm if}\;\;\delta=1,\end{array}\right.\;\;\; r=|{\vec r}|. &
\end{eqnarray}
Using equation (\ref{eq5}) and hypothesis H3, then equation (\ref{eq32}) can be written as
\begin{equation}
\label{eq32a}
\langle U^2\rangle=\int_{R^d}u^2(|{\vec r}|)\rho_{\vec R}({\vec r})d^d{\vec r}=
\left\{
\begin{array}{cc}
\displaystyle \int_{R^d}|{\vec r}|^{2(1-\delta)}\rho_{\vec R}({\vec r})d^d{\vec r}<\infty & \;\;{\rm if}\;\;\delta<1, \\ & \\
\displaystyle \int_{R^d}(\ln |{\vec r}|)^2\rho_{\vec R}({\vec r})d^d{\vec r}<\infty & \;\;{\rm if}\;\;\delta=1,
\end{array}\right.\, 
\end{equation}
and from Observation \ref{obs1} of Theorem \ref{teo1}, we have that $\alpha'>2$  for $\delta\leq 1$.

Thus, we may  apply the Theorem \ref{teo1} to obtain the following asymptotic expressions for $\Psi_{U}(b_{N}{z})^{N}$  when $N\rightarrow \infty$:
\begin{equation}
\label{eq18teo4a}
\ln\Psi_{U}(b_{N}{z})^{N}\asymp
\left\{\hspace*{2mm}
\begin{array}{ll}
\hspace*{-2mm}\displaystyle -\frac{\lambda'_{\alpha'}\Omega_d\rho_{\vec R}(\,\vec 0\,)K'}{\delta-1}\frac{|z|^{\alpha'}}{\Gamma(\alpha' +1)}\left(1-iq'\frac{z}{|z|}\tan\left(\frac{\alpha'\pi}{2}\right)\right) & \;\;{\rm if}\;\;0<\alpha'<1,
\\ \\
\hspace*{-2mm}\displaystyle{\rm i}z \left(-q'\frac{\Omega_d\rho_{\vec R}(\,\vec 0\,)K'}{\delta-1}\right)- \sigma'_N\frac{\pi}{2} \frac{\Omega_d\rho_{\vec R}(\,\vec 0\,)}{\delta-1}\frac{| z|}{\Gamma(\alpha'+1)} & \;\;{\rm if}\;\;\alpha'=1
\\ \\
\hspace*{-2mm}\displaystyle{\rm i} z\langle q'K'U\rangle-{\sigma'}_N^{\alpha'}\frac{\lambda'_{\alpha'}\Omega_d\rho_{\vec R}(\,\vec 0\,)}{\delta-1}
\frac{|z|^{\alpha'} }{\Gamma(\alpha' +1)}\left(1-iq'\frac{z}{|z|}\tan\left(\frac{\alpha'\pi}{2}\right)\right) & \;\;{\rm if}\;\;1<\alpha'< 2,
\\ \\
\hspace*{-2mm}\displaystyle{\rm i} z\langle q'K'U\rangle-{\sigma'}_N^2\frac{1}{2}\frac{\Omega_d\rho_{\vec R}(\,\vec 0\,)}{\delta-1}\frac {|z|^{2} }{\Gamma(\alpha'+1)} & \;\;{\rm if}\;\;\alpha'= 2,
\\ \\
\hspace*{-2mm}\displaystyle{\rm i}z \langle{q'K'U}\rangle-{\sigma'}_N^2\frac{\sigma^2_U z^2}{2} & \;\;{\rm if}\;\;2<\alpha'\leq\infty,
\end{array}\right.
\end{equation}
where we have used the expressions for $A_{\alpha'}(\hat z)$, $B_{\alpha'}(\hat z)$ and $G_{\alpha'}(\hat z)$, respectively given in equations (\ref{eq35}) and (\ref{eq36}). Also, let us remember that $\sigma_U^2=\left<U^2\right>$.

Now, tacking into account eq. (\ref{eq18teo4a}) and the  characteristic function of $U_{ ren}^N$ given in eq. (\ref{eq17}), we can write for the different {ranges} of $\alpha'$:
\begin{equation}
\label{eq18teo4b}
\ln\Psi_{U_{ ren}^N}(z)\asymp
\left\{\hspace*{2mm}
\begin{array}{lc}
\hspace*{-2mm}\displaystyle -\frac{\lambda'_{\alpha'}\Omega_d\rho_{\vec R}(\,\vec 0\,)K'}{\delta-1}\frac{|z|^{\alpha'}}{\Gamma(\alpha' +1)}\left(1-iq'\frac{z}{|z|}\tan\left(\frac{\alpha'\pi}{2}\right)\right) & \;\;{\rm if}\;\;0<\alpha'<1,
\\ \\
\hspace*{-2mm}\displaystyle{\rm i}z \left(-q'\frac{\Omega_d\rho_{\vec R}(\,\vec 0\,)K'}{\delta-1}\right)- \sigma'_N\frac{\lambda'_1\Omega_d\rho_{\vec R}(\,\vec 0\,)}{\delta-1}\frac{| z|}{\Gamma(\alpha'+1)} & \;\;{\rm if}\;\;\alpha'=1,
\\ \\
\hspace*{-2mm}\displaystyle{\rm i} z\langle q'K'U\rangle-{\sigma'}_N^{\alpha'}\frac{\lambda'_{\alpha'}\Omega_d\rho_{\vec R}(\,\vec 0\,)}{\delta-1}
\frac{|z|^{\alpha'} }{\Gamma(\alpha' +1)}\left(1-iq'\frac{z}{|z|}\tan\left(\frac{\alpha'\pi}{2}\right)\right) & \;\;{\rm if}\;\;1<\alpha'< 2,
\\ \\
\hspace*{-2mm}\displaystyle{\rm i} z\langle q'K'U\rangle-{\sigma'}_N^2\frac{\lambda'_2\Omega_d\rho_{\vec R}(\,\vec 0\,)}{\delta-1}\frac {|z|^{2} }{\Gamma(\alpha'+1)} & \;\;{\rm if}\;\;\alpha'= 2,
\\ \\
\hspace*{-2mm}\displaystyle{\rm i}z \langle{q'K'U}\rangle-{\sigma'}_N^2\frac{\sigma^2_U z^2}{2} & \hspace*{-10mm}\;\;{\rm if}\;\;2<\alpha'\leq\infty\;(\delta\neq 1),
\\ \\
\hspace*{-2mm}\displaystyle{\rm i}z\left(-q'K'\ln L_N+\langle{q'K'U}\rangle\right)-{\sigma'}_N^2\frac{\sigma^2_U z^2}{2} & \hspace*{-10mm}\;\;{\rm if}\;\;2<\alpha'\leq\infty\;(\delta= 1).
\end{array}\right.
\end{equation}
where we define
\begin{equation}
\label{new6}
\lambda'_{\alpha'}=\pi\frac{\cos(\alpha'\pi/2)}{\sin(\alpha'\pi)}\;\;{\rm if}\;\;0<\alpha'<2,\;\;\;\;\lambda'_2=\frac{1}{2}.
\end{equation}
Thus, considering the characteristic functions of the variables $S^*_{\alpha'}$ for $0<\alpha'<2$ ($\alpha'\neq 1$), $S_{\alpha'}$ for $\alpha'=1,2$ and $G_U$ for $\alpha'>2$, we can obtain
from eq. (\ref{eq18teo4b}) the asymptotic expressions for $U_{ ren}^N$ in eq. (\ref{eq34}). Finally, tacking the limit $N\rightarrow\infty$ in eq. (\ref{eq18teo4b}) and considering that $\lim_{N\rightarrow\infty}\sigma'_N=0$, we obtain eq. (\ref{eq34b}). {\it QED}

The analysis of the results presented in Theorem \ref{teo4} is entirely analogous to the one made about Theorem \ref{teo3}. The potential energy converges to a $\alpha'$-stable random variable in the interval $\delta>d+1$ ($\alpha'<1$) and can only be described in this limit as a strictly stochastic variable. However, unlike the force case, a stable asymmetric probability density with an asymmetric parameter given by $q'$ must describe the potential energy. On the other hand, in the interval $\delta<d+1$ ($\alpha'>1$), the potential energy converges to its average value fluctuating stochastically around it, where the width of this fluctuation tends to zero as the number of sources $N$ grows indefinitely. In the case $\delta=d+1$ ($\alpha'=1$), even if it is impossible to calculate the potential energy's average, it converges to a well-determined value. Finally, it is worth noting that for $\delta=1$, the limit of the potential energy also depends on the limit for the renormalization size $L_N$.

\subsection{The Renormalization Constants}

In this subsection, we analyze the renormalization conditions and their implications on the choices of renormalization constants $k_N$ and $L_N$ given in Definition \ref{def1}.
From a purely mathematical point of view, the constants $K$ and $K'$, respectively defined in the Theorems \ref{teo3} and \ref{teo4}, are strictly positive. From a physics point of view, the coupling constant is not arbitrary and depends on parameters with empirical meaning. Taking this into account is necessary for analyzing the sense of the possible restrictions on the choice of coupling constants and size renormalization sequences. In section 5, we illustrate this issue through the example of gravitational force.

We present in Table \ref{tabela5} below a synthetic summary of the renormalization conditions obtained in Theorems 3 and 4.
\begin{table}[h]
\begin{center}
\begin{tabular}{ccccc}
\hline\hline
$\delta$ & $\alpha$ & $\alpha'$  & $K$ & $K$' \\\hline\hline
$\delta>d+1$ & $\displaystyle \alpha=\frac{d}{\delta}<1$ &  $\displaystyle \alpha'=\frac{d}{\delta-1}<1$ &  $\displaystyle N|a_N|^\alpha$ 
& $\displaystyle N|b_N|^{\alpha'}$\\ \\
$\delta=d+1$ & $\displaystyle \alpha=\frac{d}{d+1}<1$ & $\alpha'=1$  & $\displaystyle N|a_N|^\alpha$ 
& $\displaystyle -N|b_N|\ln|b_N|$ \\ \\
$d<\delta<d+1$ &  $\displaystyle \alpha=\frac{d}{\delta}<1$ & $\alpha'>1$ &  $\displaystyle N|a_N|^\alpha$  
& $\displaystyle N|b_N|$ \\ \\
$\delta=d$ & $\alpha=1$ & $\alpha'>1$  &  $\displaystyle -N|a_N|\ln|a_N|$ & $\displaystyle N|b_N|$ \\ \\
$\delta<d$ & $\alpha>1$ & $\alpha'>1$ & $\displaystyle N|a_N|$ & $\displaystyle N|b_N|$ \\\hline\hline
\end{tabular}
\end{center}
\vspace*{-4mm}
\caption{The first column presents the different ranges for $\delta$, the second  the value of $\alpha$, the third the value of $\alpha'$, the forth and fifth column the respective renormalization conditions.}
\label{tabela5}
\end{table}

\noindent
Let us remember that $K$ and $K'$ are strictly positive constants, and the values of $a_N$ and $b_N$, respectively defined in (\ref{eq14}) and (\ref{eq16}),  are given by
\begin{equation}
\label{eq40}
a_N=\frac{k_N}{L_N^\delta},\;\;b_N=\beta a_NL_N=\frac{\beta k_NL_N}{L_N^\delta},\;\;
\beta=\left\{
\begin{array}{cc} \displaystyle \frac{1}{\delta-1} & \mbox{if}\;\;\delta\neq 1 \\
1  & \mbox{if}\;\;\delta=1
\end{array}\right. . 
\end{equation}
Moreover,  we remember that 
\begin{equation}
\label{eq41}
q=\frac{a_N}{|a_N|}=\mbox{sign} (k_N),\;\;
q'=\frac{b_N}{|b_N|}=\mbox{sign}(\beta k_N).
\end{equation}

Thus, the definitions of $a_N$ and $b_N$, the definitions of $q$ and $q'$,  and the  relations for $ K$ and $K'$ in Table V, allow us to obtain the renormalization constants $k_N$ and $L_N$. The characteristic function limits for the force and potential energy (considering the different intervals for $\delta$) are determined respectively in Theorem \ref{teo3} and Theorem \ref{teo4}.

It is worth noting that to determine the  expressions for the singular cases in Table V, we must take into account the definition for the function $h(x)$  in Theorem \ref{teo1}.
This function is defined as the unique solution of the equation
\begin{equation}
\label{funcaoh}
-h(x)\ln h(x)=x\;\;(0< x\leq e^{-1}),
\end{equation} 
where $h(x)\in (0,e^{-1}]$. The limit of $h(x)$ when $x\rightarrow 0$ can be obtained as
\begin{equation}
\label{lim_funcaoh}
\lim_{x\rightarrow 0}h(x)=-\lim_{x\rightarrow 0}\left(\frac{x}{\ln h(x)}\right)=0.
\end{equation}

In the following, we determine the expressions for $L_N$ and $k_N$ for the five intervals (Table V) of $\delta$.

\begin{renorm} {\rm (The Mean Field Limit $-$ $\delta<d$)}
\label{ren1}
The renormalization constants  for $\delta<d$ are given by:
\begin{equation}
\label{eq_ren1}
L_N=L=\frac{K'}{|\beta|K},\;\;\;\; |k_N|=\frac{KL^\delta}{N}.
\end{equation}.
\end{renorm}
\noindent {\it Proof:}
If we consider the case $\delta<d$ in Table V, then the renormalization constants should satisfy the following  equations:
\begin{equation}
\label{eq42}
\frac{N|k_N|}{L_N^\delta}= K,\;\;\;\;\frac{N|\beta||k_N|L_N}{L_N^\delta}= K', 
\end{equation} 
where $\beta$ is given in equation (\ref{eq40}). Solving equations in (\ref{eq42}) with respect to the renormalization constants, we obtain:
\begin{equation}
\label{eq43}
L_N= L=\frac{K'}{|\beta|K},\;\;\;\;|k_N|=\frac{{K'}^\delta K^{1-\delta}}{|\beta|^\delta}\frac{1}{N}. 
\end{equation}
Finally, solving  equation for $L$  with respect to $K'$ and substituting into the  equation for $k_N$  in (\ref{eq43}), we obtain $|k_N|=KL^\delta/N$.
{\it QED}\\

We see that $L_N$ does not depend on $N$, which shows that the system size must keep constant in the renormalization process.
The coupling $k_N$  scales like $N^{-1}$. It is typical of Kac scaling in Mean Field Theory \cite{Campa2009, Kac}.
Moreover, we have $\lim_{N\rightarrow\infty}|k_N|=0$,
which characterizes the so-called weak coupling, that is, the renormalized coupling constant  $k_N$ tends to zero as the number of sources tends to infinity.

Let us observe that for $\delta=1$, we must consider $\lim_{N\rightarrow\infty}L_N=L$ in the expression for the characteristic function in Theorem \ref{teo4}.

\begin{renorm} {\rm (The Thermodynamic Limit $-$ $\delta>d+1$)}
\label{ren2}
If $\delta>d+1$, then the renormalization constants satisfy the following conditions:
\begin{equation}
\label{eq_ren2}
L_N= (\delta-1)\left(\frac{{K'}^{\delta-1}}{K^\delta}\right)^{1/d}N^{1/d},
\;\;\;\;|k_N|=|k|=(\delta-1)^\delta\left(\frac{K'}{K}\right)^{\delta(\delta-1)/d}.
\end{equation}
\end{renorm}
\noindent{\it Proof:}
For $\delta>d+1$ in Table V, we have:
\begin{equation}
\label{eq48}
N\left(\frac{|k_N|}{L_N^\delta}\right)^{d/\delta}= K,\;\;\;\;N\left(\frac{|k_N|L_N}{( \delta-1)L_N^\delta}\right)^{d/(\delta-1)}= K',
\end{equation}
where  $\beta=1/(\delta-1)$ in eq. (\ref{eq40})  since $\delta>1$. Solving equation (\ref{eq48}) with respect to $L_N$ and $k_N$, we obtain (\ref{eq_ren2}).
{\it QED} \\

From (\ref{eq_ren2}), we see that the relationship between $L_N$ and $N$ is given by
\begin{equation}
\label{eq50}
\frac{N}{L_N^d}= \rho=\frac{1}{(\delta-1)^d}\frac{K^{ \delta}}{{K'}^{\delta-1}},
\end{equation}
where $\rho$ is a strictly positive constant. This is typical of the so-called Thermodynamic Limit:
the system size given by $L_N$ tends to infinity as $N$ tends to infinity in such a way that the system volume  (which is proportional to $L_N^d$) grows linearly with $N$.

Another important property is that $k_N$ must not be renormalized and held constant through the renormalization process.
We call this kind of renormalization as strong coupling, where the renormalized coupling $k_N$ does not converge to zero as $N$ grows to infinity.

Finally, let us remark that if $\rho$ and $k$ are given, then $K$ and $K'$ must be defined as:
\begin{equation}
\label{eq52}
K=\rho |k|^{d/\delta},\;\;\;\;K'=\rho\left(\frac{|k|}{\delta-1}\right)^{d/(\delta-1)}.
\end{equation}

\begin{renorm} {\rm (The Mixed Limit $-$ $d< \delta < d+1$)}
\label{ren3}
Considering the range $d< \delta < d+1$,  the renormalization constants are given by:
\begin{equation}
\label{eq_ren3}
N\rightarrow\infty\;\;\Rightarrow\;\;
L_N=\frac{(\delta-1)K'}{K^{\delta/d}}N^{(\delta-d)/d},\;\;\;\;|k_N|=\left( \frac{(\delta-1)K'}{K^{(\delta-1)/d}}\right)^{\delta}\frac{1}{N^{\delta(d+1- \delta)/d}}
\end{equation}
\end{renorm}
\noindent{\it Proof:} For $d<\delta<d+1$ in Table V, we have the following  renormalization relations:
\begin{equation}
\label{eq53}
N\left(\frac{|k_N|}{L_N^\delta}\right)^{d/\delta}= K,\;\;\;\;N\left(\frac{|k_N|L_N}{( \delta-1)L_N^\delta}\right)= K',
\end{equation}
where again from equation (\ref{eq40}), we have used $\beta=1/(\delta-1)$ since $\delta>1$. Solving  equation (\ref{eq53}) for $L_N$ and $k_N$, we obtain the  expressions for the
renormalization constants. {\it QED} \\

From (\ref{eq_ren3}), we have the following relationship between $L_N$ and $N$:
\begin{equation}
\label{eq_ren3a}
\frac{N^{\delta-d}}{L_N^d}=\nu=\left(\frac{K^{\delta/d}}{(\delta-1)K'}\right)^d,
\end{equation}
where $\nu$ is a strictly positive constant. Also, we can see that
\begin{equation}
\label{eq55}
\lim_{N\rightarrow\infty}L_N=\infty,\;\;\;\;\lim_{N\rightarrow\infty}{|k_N|}=0.
\end{equation}
In this case, the  behavior for the renormalization constants has something in common with both limits: the Mean Field Limit ($\delta<d$) and the Thermodynamic Limit $(\delta>d+1$).

As in the Mean Field limit, the coupling constant $k_N$ must go to zero as $N$ goes to infinite, while, as in the thermodynamic limit, the spatial size of the source system must go to infinity. However, in both cases, the  dependence on $N$ does not follow the same functional form as the respective Mean Field and Thermodynamic Limit.
Indeed, from relations given in Renormalization \ref{ren3}, we see that  $L_N$ grows  slower than in the thermodynamic case,
just as $k_N$ tends to zero slower than in the Vlasov limit.

\begin{renorm} {\rm (The Singular Limit $-$ $\delta=d$)}
\label{ren4}
Considering  $\delta=d$, the renormalization constants are given by:
\begin{equation}
\label{eq_ren4}
N\rightarrow\infty\;\;\Rightarrow\;\;
L_N=\frac{K'}{|\beta|}\frac{1}{Nh(K/N)},\;\;\;\;|k_N|=\left(\frac{K'}{|\beta|}\right)^d\frac{1}{N^d[h(K/N)]^{d-1}},
\end{equation}
where the function $h(K/N)$ is  defined in (\ref{funcaoh}).
\end{renorm}
\noindent {\it Proof:} For $\delta=d$ the  renormalization equations in Table V imply that
\begin{equation}
\label{eq56}
|a_N|= h\left(\frac{K}{N}\right)=\frac{|k_N|}{L_N^d},\;\;\;\;N\frac{|\beta||k_N| L_N}{L_N^d}=K',
\end{equation}
where the function $h(K/N)$ is defined in (\ref{funcaoh}) and represents the solution of the renormalization equation for $|a_N|$ in Table V. Here, we must take into account the value of
$\beta$ in (\ref{eq40}) according to the value of $\delta=d$ considered.
Solving (\ref{eq56}) for $L_N$ and $k_N$, we get the  expressions for the renormalization constants. {\it QED}\\

From the properties of the function $h(x)$ when $x\rightarrow 0$, we can show that:
\begin{equation}
\label{eq58}
\lim_{N\rightarrow\infty}L_N=\infty,\;\;\;\;\lim_{N\rightarrow\infty}{|k_N|}=0.
\end{equation}

\begin{renorm} {\rm (The Singular Limit $\delta=d+1$)}
\label{ren5}
Considering $\delta=d+1$,  the  expressions for the renormalization constants are given by:
\begin{equation}
\label{eq_ren5}
N\rightarrow\infty\;\;\Rightarrow\;\;
L_N=\left(\frac{d}{K^{(d+1)/d}}\right)h\left(\frac{K'}{N}\right)N^{(d+1) /d},\;\;\;\;|k_N|=\left(\frac{d}{K}\right)^{d+1}\left[N h\left(\frac{K'}{N }\right)\right]^{d+1},
\end{equation}
where  the function $h(K/N)$ is defined in (\ref{funcaoh}).
\end{renorm}
\noindent{\it Proof:}
For $\delta=d+1$ in Table V, we have the following  renormalization equations:
\begin{equation}
\label{eq59}
N\left(\frac{|k_N|}{L_N^{d+1}}\right)^{d/(d+1)}= K,\;\;\;\;|b_N|= h\left( \frac{K'}{N}\right)=\frac{|k_N|L_N}{dL_N^{d+1}},
\end{equation}
where $\beta=1/(\delta-1)=1/d$ in (\ref{eq40}) and $h(K'/N)$ is the solution of the renormalization equation for $|b_N|$.
Solving  equation (\ref{eq59}) for $L_N$ and $k_N$, we obtain the  renormalization constants. {\it QED}\\

Using the properties of the $h(x)$ function, we can show that the coupling is weak and  the size of the system must grow indefinitely with the increase of $N$, that is,
\begin{equation}
\label{eq60}
\lim_{N\rightarrow\infty}L_N=\infty,\;\;\;\;\lim_{N\rightarrow\infty}{|k_N|}=0. 
\end{equation}

\section{Discussion and Analysis of the Renormalization Limits}

The first issue is to analyze how the source's probability distribution determines the type of stable Lévy distribution, which characterizes the force and potential energy on the test particle. The Theorem \ref{teo3}, which fixes the exponent  $\alpha$ for the characteristic function of the force, { assumes} only H2. The Theorem \ref{teo4}, which determines the exponent $\alpha'$ for the characteristic function of the potential energy,  { assumes}  H2 for $\delta> 1$ and H3 for $\delta\leq 1$. We show that $\alpha=d/\delta$ for  $\delta>0$, $\alpha'=d/(\delta-1)$ for $\delta>1$ and $\alpha'>2$ for $\delta\leq1$.

It is worth noting that the only supposition in H2 is the continuity of the source's { measure} at any possible position of the test particle. If H2 is not fulfilled, it is possible to show (see reference \cite{Figueiredo2019}) that the exponents $\alpha$ and $\alpha'$ do not change. However, the characteristic function for the force would correspond to an asymmetric Lévy distribution. For the case where H3 is not satisfied, we can not assure $\alpha'>2$.

Another important issue is whether or not to use a random vector or a random variable to respectively describe the force or the potential energy when $N\rightarrow\infty$. In this aspect, we can see that the Mean Field Limit and  Thermodynamic Limit represent opposite situations, while the mixed limit represents an intermediate situation.

In the Mean Field Limit, force and potential energy converge to fixed values determined by their respective averages. Indeed, the stochastic nature of the model disappears when $N\rightarrow\infty$. In the Thermodynamic Limit, force and potential energy are determined respectively by a random vector and random variable when $N\rightarrow\infty$. Finally, the mixed limit presents an intermediate situation: the potential energy converges to its average, while a random vector determines the force.

Let us now analyze the issue of the relationship between the mean force and mean potential energy in the Mean Field Limit (Renormalization \ref{ren1} for $\delta<d$).
In order to proceed, let us consider a system of units with length unit $l_0$ and coupling constant unit $k_0$ - see equation (\ref{forca_dimensao}). Also, let us consider a random vector $\vec R$ as established in Definition \ref{def1}.  Taking into account that ${\vec R}={\vec y}-{\vec Y}$ and the Theorem \ref{teo3} for $\delta<d$, the limit for the  force can be written as:
\begin{equation}
\label{eq63}
{\vec F}_{ ren}=\langle qK{\vec V}\rangle= \mbox{sign}(k_N)K\int_{\mathbb{R}^d}\frac{{\vec y}-{\vec y}_i}{\left|{\vec y}-{\vec y }_i\right|^{\delta+1}}\rho_{\vec Y}({\vec y}_i)d^d{\vec y}_i,
\end{equation}

From Theorem \ref{teo4} for $\delta<d$ and the relation $K'=KL$ in Renormalization \ref{ren1}, the potential energy can be written as:
\begin{equation}
\label{eq64}
U_{ ren}({\vec y})=\langle q'K' U\rangle=
\left\{\begin{array}{cc}
\displaystyle \frac{\mbox{sign}((\delta-1)k_N)}{|\delta-1|}KL\int_{R^d}|{\vec y}-{\vec y}_i| ^{(1-\delta)}\rho_{\vec Y}({\vec y}_i)d^d{\vec y}_i & \;\;{\rm if}\;\;\delta\neq 1,\\ & \\
\displaystyle \mbox{sign}(k_N)KL\int_{R^d}\ln {|{\vec y}-{\vec y}_i|}\rho_{\vec Y}({\vec y}_i)d ^d{\vec y}_i & \;\;{\rm if}\;\;\delta=1
\end{array}\right.\, 
\end{equation}

Some remarks are relevant here. Hypothesis H1 follows from the supposition that the sources have the same spatial distribution, meaning the same random vector ${\vec Y}$ generates their positions. It implies that the relative position distribution of the test particle concerning each source is also described by the same random vector ${\vec R}$. However, for each fixed position ${\vec y}$ of the test particle, the random vector ${\vec R}$ is not necessarily the same. A consequence of this property can be seen in equations (\ref{eq63}) and (\ref{eq64}), where the force and potential energy may depend on ${\vec y}$.
 
The only form to define a  potential energy such that ${\vec F}_{ ren}({\vec y})=-\nabla_{\vec y} U_{ ren}({\vec y})$ is to consider that $L_N=1$ in the renormalization process, that is, the sources distribution must be kept unchanged and the renormalized coupling defined as $|k_N|=K/N$ for a given $K>0$. If we define $\kappa={\rm sign}(k_N)K$, then we have
\begin{equation}
\label{eq67}
-\nabla_{\vec y} U_{ ren}({\vec y})=
\left\{\begin{array}{cc}
\displaystyle \frac{\kappa}{\delta-1}\int_{R^d}-\nabla_{\vec y}\left(|{\vec y}-{\vec y}_i|^{1-\delta}\right)\rho_{\vec Y}({\vec y}_i)d^d{\vec y}_i & (\delta\neq 1) \\ & \\
\displaystyle -\kappa\int_{R^d}-\nabla_{\vec y}\left(\ln |{\vec y}-{\vec y}_i|\right)\rho_{\vec Y}({\vec y}_i)d^d{\vec y}_i & (\delta=1).
\end{array}\right\}={\vec F}_{ ren}({\vec y})\, .
\end{equation}

In a completely different way, the Thermodynamic Limit  (Renormalization 2 - $\delta>d+1$) does not allow to keep, in the renormalization process, the size of the source system unchanged since $\lim_{N\rightarrow\infty}L_N=\infty$. Therefore, the renormalization implies that the system size $L_N$  must grow with the number of sources $N$.
Considering the Theorems 3 and 4 respectively for $\alpha<1$ and $\alpha'<1$, we can write: 
\begin{eqnarray}
{\vec F_{ ren}}&=&\sigma_{\alpha}({\vec y}){\vec S}_{\alpha}\;\;{\rm for}\;\;\alpha=\frac{d}{\delta}<1, \nonumber \\
{U^N_{ ren}}&=&\sigma'_{\alpha'}({\vec y}){S}^*_{\alpha'},\;\;{\rm for}\;\;\alpha'=\frac{d}{\delta-1}<1, \label{eq69}
\end{eqnarray}
where ${\vec S}_\alpha$ is { an} $\alpha$-stable symmetric random vector with exponent $\alpha<1$,  $S^*_{\alpha'}$ is a $\alpha'$-stable asymmetric random variable with exponent $\alpha'<1$ and asymmetry parameter given by $q'=\mbox{sign}(k_N)$. The functions $\sigma_{\alpha}({\vec y})$ and $\sigma'_{\alpha'}({\vec y})$ are respectively given by:
\begin{eqnarray}
\label{larguras}
\sigma_\alpha(\vec y)&=&\left(\frac{\lambda_\alpha K\rho_{\vec R}(\,\vec 0\,)}{\delta\Gamma(\alpha+1)}\right)^{1/\alpha}=\left(\frac{\lambda_\alpha K\rho_{\vec Y}({\vec y})}{\delta\Gamma(\alpha+1)}\right)^{1/\alpha},\nonumber\\
\sigma_\alpha'(\vec y)&=&\left(\frac{\lambda'_{\alpha'}\Omega_dK'\rho_{\vec R}(\,\vec 0\,)}{(\delta-1)\Gamma(\alpha'-1)}\right)^{1/\alpha'}=
\left(\frac{\lambda_{\alpha'}\Omega_dK'\rho_{\vec Y}({\vec y})}{(\delta-1)\Gamma(\alpha'-1)}\right)^{1/\alpha'}.
\end{eqnarray}

Therefore, we see that the description of the limits for the force and potential energy requires random quantities with stable Lévy distributions. There is no sense in defining mean quantities for the force and potential energy since the Lévy exponents are less than one. Let us observe that $\sigma_\alpha(\vec y)$ and $\sigma'_{\alpha'}(\vec y)$, which respectively measure the dispersion width for the force and potential energy, depending on the test particle position.

The mixed case (Renormalization \ref{ren3} - $d<\delta<d+1$) combines the Mean Field and Thermodynamic Limits,
with a well-defined mean for potential energy but no well-defined mean for force.
For this case, the force is linked to a random stable Lévy vector, and the former is given by equation (\ref{eq64}) for $\delta\neq 1$. Due to the impossibility of commuting the differentiation operator with the integration operator in  (\ref{eq67}), the force can not be calculated as $-\nabla_{\vec y}(U_{ ren})$ and $L_N$  { diverges}. However, like the Mean Field Limit, the coupling constant $k_N$ must tend to zero when $N\rightarrow\infty$.

Concerning  behavior for the renormalization constants $L_N$ and $k_N$, singular and mixed cases are similar. However, from the point of view of the existence of non-random quantities that determine the force and the potential energy, the case $\delta=d$ is similar to the Mean Field Limit. While the case involving $\delta=d+1$ is comparable to the Thermodynamic Limit.

Finally, let us stress that no assumption about global isotropy or uniformity was made for the random vector ${\vec Y}$. The hypothesis H2 { assumes} only that the probability density $\rho_{\vec Y}({\vec y}_i)$ (for each source $i=1,\ldots,N$) is continuous at any test particle position ${\vec y}_i={\vec y}$. Mathematically, it { consists in}  considering that the { measure} $\rho_{\vec Y}({\vec y}_i)$ is a continuous function defined on $\mathbb{R}^d$.
One can interpret it as a local isotropy and uniformity hypothesis for the sources. In other words, for a sufficiently small neighborhood of a test particle, the source particles inside this neighborhood have an approximately uniform and isotropic distribution.

\subsection{On the determination of $K$ and $K'$}

The meaning of strong or weak couplings is purely mathematical. Indeed, we only want to characterize whether the force constant $k_N$ should be renormalized or not, depending on the number of sources. Applied to a physical problem, this means whether or not to change the force intensity to obtain well-determined statistical limits to the force and potential energy. The renormalizations discussed in section 5 establish the mathematical conditions for assuring a well-defined probability distribution for many sources. Whether or not these mathematical conditions are satisfied or have a meaning in physical systems, it is essential to analyze every specific problem to answer these questions.

Let us look at the strong coupling case associated with the Thermodynamic Limit for $\delta>d+1$, where the force parameter $k_N$ must be kept constant in the renormalization process. Such a mathematical requirement has no physical implications in the coupling $k$ described by equation (\ref{eq10}). However, this renormalization implies that the spatial size of the source system needs to increase while preserving the particle density $\rho$ defined in (\ref{eq50}).
Therefore, once given a coupling $k$ and density $\rho$ the constants $K$ and $K'$ are obtained through eq. (\ref{eq52})
As a physical consequence, we have the force and potential energy described by stable Lévy probability densities with dispersion widths dependent on $\rho$ and $k$ since they depend on $K$ and $K'$ as shown in eqs. (\ref{larguras}).

For weak coupling, $k_N$ must depend on $N$, and its value must tend to zero when $N\rightarrow\infty$, as obtained in all renormalizations except the one corresponding to the Thermodynamic Limit (Renormalization \ref{ren2}). The term weak coupling stems from this convergence to zero. The interpretation and meaning of this type of renormalization can only be assessed in terms of the physical parameter (from which the coupling is defined) that is picked to be renormalized.

We will illustrate the meaning of a weak coupling through an example of $N$ sources of the same mass interacting through a gravitational force with the test particle.
In this example, we have $\delta=2$ and $k=-Gm_{t}m$,
where $m$ denotes the mass of each source, $m_t$ is the mass of the test particle, and $G$ is the universal constant of gravitation. Let us remark that these parameters are all determined in a given system of units. The sign of $k$ is negative since the force is attractive and must be pointing from the test particle to the source. Let us pick the source mass as a parameter to be renormalized and, therefore, dependent on $N$. Denoting the source renormalized mass by $m_N$, the renormalized coupling is given by $k_N=-Gm_tm_N$. Also, from eqs. (\ref{eq40}) and (\ref{eq41}), we have that $\beta=1$ and $q'=-1$ . Let us now analyze the three cases corresponding to the different spatial dimensions $d=3$, $d=2$, and $d=1$.

\subsubsection{Case $d=3$ and $\delta=2$}

In this case, we have $\delta<d$ and we must apply the results of the Renormalization \ref{ren1}:
\begin{equation}
\label{grav1}
L_N= 1,\;\;  K= Nm_tm_N,\;\;   K'= Nm_tm_N
\end{equation}
Considering the total mass of the source system given by $M_N=Nm_N$ and substituting in to equation (\ref{grav1}), we obtain
\begin{equation}
\label{grav2}
K= m_t M_N,\;\;K'=m_t M_N.
\end{equation}
Since $K$ and $K'$ are constant then the total mass of the source system $M_N=M$ must be constant as well. Therefore, the renormalized mass $m_N$ and the respective renormalization constants must be defined as
\begin{equation}
\label{grav3}
m_N=\frac{M}{N},\;\;K=K' =m_tM.
\end{equation}
The interpretation of this renormalization is clear: the size and mass of the source system remain constant as the number of sources increases.
The constants $K$ and $K'$ are not arbitrarily chosen. The masses of the test particle and source system (equation (\ref{grav3})) determine their values.
The exponents of Lévy are respectively $\alpha=3/2$ and $\alpha'=3$, and both the force and potential energy converge to their means when the number of sources grows indefinitely (see Theorems \ref{teo3} and \ref{teo4}).

The asymptotic stochastic fluctuations around these means are described respectively by { an} $\alpha$-stable Lévy probability density with $\alpha=3/2$ for the force (see Theorem \ref{teo3}) and by a Gaussian probability density ($\alpha'>2$) for the potential energy (see Theorem \ref{teo4}). In both cases, the widths of these fluctuations decrease to zero as the number of sources increases indefinitely.

As an historical remark, we have that the Lévy's $\alpha$-stable density for $\alpha=3/2$ was originally described by Holtsmark \cite{hot-1919} in the analogous problem of a Coulomb electrostatic force. 

\subsubsection{$d=2$ and $\delta=2$}

In this case, we have $\delta=d$ and we must apply the Renormalization 4. The relations in eq (\ref{eq56}), written in terms of the coupling $k_N=-Gm_tm_N$, become
\begin{equation}
\label{grav4}
h\left(\frac{K}{N}\right)=\frac{Gm_tm_N}{L_N^2},\;\;K' =\frac{NGm_tm_N}{L_N}.
\end{equation}

Defining the total mass of the sources as $M_N=Nm_N$ and substituting in to equation (\ref{grav4}):
\begin{equation}
\label{grav5}
K= Nh^{-1}\left(\frac{Gm_tM_N}{NL_N^2}\right),\;\;;K'=\frac{Gm_tM_N}{L_N},
\end{equation}
where { $h^{-1}=-x\ln x$ is the inverse function of $h(x)$}. We see that as $K'$ is constant then $M_N/L_N=\mu>0$ must be constant as well. Therefore, we can choose any $\mu>0$ and $K>0$ such that
\begin{equation}
\label{grav6}
K= Nh^{-1}\left(\frac{Gm_t\mu}{L_N}\right),\;\;;K'={Gm_t\mu}.
\end{equation} 
Finally, from the relations in eq. (\ref{eq_ren4})  the size and mass renormalizations are defined  as
\begin{equation}
\label{garv7}
 L_N= \frac{Gm_t\mu}{Nh\left(\displaystyle\frac{K}{N}\right)},\;\;m_N=\frac{M_N}{N}=\frac{\mu L_N}{N}=\frac{Gm_t\mu^2}{N^2\displaystyle h\left(\frac{K}{N}\right)}.
\end{equation} 

The Lévy exponents are given by $\alpha=1$ and $\alpha'=2$, showing that the force (even without a well-defined mean) converges to zero (see Theorem \ref{teo3}) and the potential energy converges to its average value (see Theorem \ref{teo4}). In both cases, the asymptotic stochastic fluctuations decrease with the increasing number of particles. With exponents $\alpha=1$ for the force and $\alpha'=2$ for the energy, symmetric and stable Lévy probability densities describe the cases.

\subsubsection{$d=1$ and $\delta=2$}

In this case, as $\delta=d+1$, we have to use the Renormalization \ref{ren5} and the relations in eq. (\ref{eq59}), which written in terms of the constant $k_N=-Gm_tm_N$ become
\begin{equation}
\label{grav8}
\frac{N(Gm_tm_N)^{1/2}}{L_N}= K,\;\;h\left(\frac{K'}{N}\right)=\frac{Gm_tm_N}{L_N}.
\end{equation}
Considering the total mass of sources as $M_N=Nm_N$, the relations in (\ref{grav8}) can be rewritten as:
\begin{equation}
\label{grav9}
Gm_t\frac{NM_N}{L^2_N}= K^2,\;\;K'=Nh^{-1}\left(\frac{Gm_tM_N}{NL_N}\right).
\end{equation}
The fact that $K$ is constant implies that $\nu=NM_N/L_N^2$ must be constant. Solving  equation for $K$ with respect to $Gm_tM_N$ and substituting in to equation for $K'$ in (\ref{grav9}), we obtain
\begin{equation}
\label{grav10}
Gm_t\nu= K^2,\;\;K'=Nh^{-1}\left(\frac{K^2L_N}{N^2}\right)
\end{equation}
Finally, the size and mass renormalizations are defined as
\begin{equation}
\label{grav11}
L_N=\frac{1}{K ^2}N^2h\left(\frac{K'}{N}\right),\;\;m_N=\frac{M_N}{N}=\frac{\nu L_N^2}{N^2}=\frac{\nu}{K^4}\left[Nh\left(\frac{K'}{N}\right)\right]^2
\end{equation}
It is evident that we can choose any value $\mu>0$ and $K'>0$.

\section{Concluding Remarks}

This work discusses the interaction between $N$ random point sources and a fixed test particle.
As a main result, we determine the renormalization conditions for assuring that the stochastic model developed in section 3 leads to a well-defined limit for the force and potential energy on the test particle when $N$ grows to infinite.
Indeed, it is impossible that these two quantities were well determined mathematically without imposing renormalization conditions in the coupling and the spatial size of the system.

We characterized five different limits and ranges for the force exponent $\delta$: the Mean Field Limit ($\delta<d$), where the spatial size remains constant, and the coupling constant decreases proportionally to $N^{-1}$; the Thermodynamic Limit ($\delta>d+1$), where the spatial size grows proportionally to $N^{1/d}$, the coupling is constant, and the mixed limit ($d\leq \delta\leq d+1$), where the coupling decreases concerning $N$  more slowly than in the Mean Filed Limit and the spatial size increases slower than in the Thermodynamic Limit. The two singular cases are comparable to mixed limits, as it preserves Mean Field Limit (decreasing coupling) and the Thermodynamic Limit (increasing spatial size) characteristics.

We show that the Central Limit formulated in the Theorem \ref{teo1} solves the problem described in section 3, where the force and potential energy probability distributions require a finite limit. In fact, from various cases of renormalization obtained in the reference \cite{Figueiredo2019}, only five types of possible renormalization remain, as described in the Renormalizations \ref{ren1}-\ref{ren5} of section 4C. All renormalizations, other than these five and which guarantee a well-defined limit for the force, will necessarily lead to divergence in the probability distribution of the potential energy.

A relevant observation concerns hypothesis H3 made for the case that the exponent of the force is $\delta\leq1$. This hypothesis guarantees that the Lévy exponent of the potential energy is $\alpha'>2$. However,
it is impossible to obtain the value of this exponent without knowing the tail behavior of the source's { measure}. In fact, for $\delta>1$, the exponent  $\alpha'$-Lévy depends only on the $\delta$-force exponent and the dimension $d$ of the space, as the tails of the energy { measure} are determined only by the behavior of the source { measure} close to the test particle.
On the other hand, for $\delta\leq 1$, the tail of the energy { measure} depends on the behavior of sources far from the test particle. Thus, a complete description for the case $\delta\leq 1$, with their respective calculation of the $\alpha'$-Lévy exponent, would require classifying all possible tails for the source's { measures}.
The $\alpha'$-Lévy exponent could assume values smaller than 2, and a complete classification for the limits would be performed in addition to Theorem \ref{teo4}. For simplicity, we use hypothesis H2 to express all limits in a single theorem for all values of $\delta$. As a proposal for future work, we suggest a study of $\delta$-force exponents in the range $\delta\leq 1$.

A generalization of this work can be made for the case where the couplings of each source are different, that is, the coupling $k$ in equation  (\ref{forca_dimensao}) can be different for each source. In this case, it is enough to adapt the theorem formulated by Lévy in reference \cite{Lévy-1924} - where the sum of a linear combination of identical variables is considered - for the case of a linear combination of random vectors. This study will be the subject of  a future work.
 
\begin{acknowledgments}
E.L.S. Silva and A. Figueiredo acknowledge CNPq for the financial support. The authors thank the fruitful comments of the Referee, and M.A. Amato, for providing valuable feedback and suggestions.
\end{acknowledgments}

\end{document}